\documentclass[
aps,11pt,
superscriptaddress,
amsmath,amssymb, nofootinbib,
prd]{revtex4-2}

\pdfoutput=1
\usepackage{ifpdf}
\usepackage[utf8]{inputenc}
\usepackage{mathdots}
\usepackage{setspace}
\usepackage{url}
\usepackage{amscd,braket,bm}
\usepackage{amsfonts}
\usepackage{color}
\usepackage{graphicx}
\usepackage{xcolor}
\usepackage{accents}
\usepackage[colorlinks=true,backref=false, linktocpage=true,
citecolor=blue,urlcolor=blue,linkcolor=blue,pdfpagemode=UseOutlines,pdfstartview=FitH,bookmarksopen]{hyperref}
\usepackage{soul}

\newcommand{\cmis}{c_\infty}

\newcommand{\be}{\begin{eqnarray}}
\newcommand{\ee}{\end{eqnarray}}

\newcommand{\ba}{\begin{align}}
\newcommand{\ea}{\end{align}}

\newcommand{\rfn}[1]{~(\ref{#1})}
\newcommand{\rfe}[1]{Eq.~(\ref{#1})}
\newcommand{\rfes}[1]{Eqs.~(\ref{#1})}
\newcommand{\rfs}[1]{Sec.~\ref{#1}}

\newcommand{\rff}[1]{Fig.~\ref{#1}}

\newcommand{\rfcs}[1]{Refs.~\cite{#1}}
\newcommand{\rfa}[1]{Appendix~\ref{#1}}

\newcommand{\f}[2]{\frac{#1}{#2}}

\newcommand{\edens}{\mathcal{E}}

\newcommand{\pa}{\mathcal{A}}

\newcommand{\perparea}{A_\perp}
\def\umu{{u^\mu}}
\def\pimunu{{\pi^{\mu \nu}}}

\newcommand{\av}[1]{\langle{#1}\rangle_\perp}

\begin{document}

\title{QGP Physics from Attractor Perturbations}

\author{Xin An}
\email{xin.an@ncbj.gov.pl}
\affiliation{National Centre for Nuclear Research, 02-093 Warsaw, Poland}

\author{Micha\l\ Spali\'nski}
\email{michal.spalinski@ncbj.gov.pl}
\affiliation{National Centre for Nuclear Research, 02-093 Warsaw, Poland}
\affiliation{Physics Department, University of Bia{\l}ystok,
  15-245 Bia\l ystok, Poland}

\begin{abstract}

    The strong longitudinal expansion characteristic of heavy-ion collisions
    leads to universal attractor behavior of the resulting drop of quark-gluon
    plasma (QGP) already at very early times.
    Assuming approximate boost invariance
    and neglecting transverse expansion at the initial time,
    we incorporate subsequent transverse dynamics of this system by linearizing the
    Mueller-Israel-Stewart  theory around the
    transversely homogeneous
    attractor.  The result is a system
    of coupled ordinary differential equations which describes the proper-time
    evolution of Fourier modes encoding the transverse structure of the initial
    energy deposition. The late-time asymptotic behavior of these solutions is
    described by transseries which make manifest the stability of the attractor
    against transverse perturbations. In this framework,
    information about the structure of the plasma in the transverse plane
    resides mostly in exponentially suppressed corrections to
    evolution along the attractor, which are not yet negligible at freeze-out.
    These findings also suggest a simple numerical approach to QGP dynamics
    using a finite number of Fourier
    modes. Physical observables can be expressed in terms of the asymptotic data
    evaluated at freeze-out. We demonstrate the efficacy of this approach by calculating the final
    multiplicity distributions and collective flow coefficients.

\end{abstract}

\maketitle

\section{Introduction}

Quark-gluon plasma (QGP) is created in heavy-ion collision experiments in highly
anisotropic, nonequilibrium states. Many features of the subsequent evolution
are successfully described by models formulated in the language of fluid
dynamics, which are applied long before local equilibrium is established. This
implies a vast reduction in the number of degrees of freedom at the earliest
moments following the collision. A possible explanation of this follows from a
key kinematical feature of heavy-ion collisions: the dominant longitudinal
expansion at the prehydrodynamic stages of evolution. An idealization of this
situation assumes boost invariance along the collision axis and neglects the
transverse dynamics~\cite{Bjorken:1982qr}. It has been shown in a number of
models that in such circumstances an early-time, far-from-equilibrium attractor
governs the dynamics until the QGP drop approaches a state amenable to a
hydrodynamic description with small
gradients~\cite{Heller:2015dha,Romatschke:2017vte,Kurkela_2020,Blaizot:2017ucy}.
Within such a picture, the information about the initial state is contained in a
single scale which characterizes the particular attractor background, up to
corrections which are exponentially suppressed -- and thus effectively lost --
at asymptotically late times.  Of course, what actually happens in heavy-ion
collision experiments is that the system does not survive until such
asymptotically late times, because as the effective local temperature drops, the
cooling QGP is converted into a stream of hadrons which source the multitude of
particles registered in detectors. This results in a wealth of information which
reflects the structure of the initial state primarily encoded in the dynamics in
the plane transverse to the collision axis.  Our goal is to extend the Bjorken
model so as to account for transverse dynamics in a way which still allows for
some analytic insights. We will show that within such a description, essentially
all of the information about the initial state encoded in the spectra of
outgoing hadrons resides in exponentially suppressed terms which are not yet
negligibly small at the time of freeze-out.
  While this picture must still be viewed as a toy model, it provides a
  managable playground which extends the range of phenomena which can be
  addressed, well beyond the scope of the original Bjorken model.

We focus on a version of the theory originally due to
Müller~\cite{Muller:1967zza} and Israel and
Stewart~\cite{Israel:1976tn,Baier:2007ix} (MIS) which can be viewed as a
possible ``UV-completion'' of Navier-Stokes (NS) theory of hydrodynamics.
It describes the
universal hydrodynamic regime which emerges at late times, but also includes a
nonhydrodynamic sector which is necessary for causality~\cite{Heller:2022ejw}
and provides a very simple model of the dynamics at earlier times (see
e.g.~\cite{Spalinski:2016fnj,Florkowski:2017olj,Jankowski:2023fdz}).  Our
analysis retains the assumption of longitudinal boost invariance, and accounts
for the transverse dynamics at the level of linearization around the
boost-invariant and transversely homogeneous attractor. At the initial time the
expansion is assumed to be purely longitudinal, but the dynamics induces a
build-up of transverse flow which expresses the structure of the energy density
at the initial time.  The linearized fields capture all the dependence of the
fireball on the transverse coordinates, and are conveniently analyzed in Fourier
space. Under this assumption, the full MIS equations reduce to a system of
coupled ordinary differential equations for a set of modes parametrized by the
transverse wave vector.

The asymptotic solutions of this system at large proper time can be studied
analytically.  Since the background is homogeneous in the transverse plane, its
dynamics carries no information about the transverse structure of the initial
state. In fact, the only information about the initial state carried by the
background solution which is not exponentially suppressed at late times is the
overall energy scale (see e.g.~\cite{Florkowski:2017olj}). All the information
about the transverse structure of the initial energy deposition is encoded in
the evolution of the perturbations.  We find that these perturbations of the
background are exponentially damped (apart from zero modes) and show that
essentially all physical observables are determined by these corrections.  Thus,
the physics of QGP flow in heavy-ion collisions provides an example of a very
nontrivial dynamical system where the exponentially damped corrections to
asymptotic results are not just non-negligible, but in fact contain almost all
of the physically relevant information.  One can usefully distinguish three
phases of evolution: the expansion-dominated initial stage whose course depends
weakly on the parameters of the theory, the asymptotic near-equilibrium
hydrodynamic stage where all information about the transverse dynamics is
exponentially suppressed and effectively undetectable, and the intermediate
stage where the transseries corrections are not yet negligible and carry almost
all the information about the initial state. They also reflect the
nonhydrodynamic content of the theory.

Linearized perturbations around relativistic hydrodynamic flows were extensively
investigated some years ago in the context of Navier-Stokes theory (see, e.g.,
Ref.~\cite{Floerchinger:2011pxy}). It has since been recognized that in the
context of heavy-ion physics one needs models which describe equilibration at
times when the nonhydrodynamic sector still plays a crucial role, which
manifests itself through the form of the attractor and its perturbations. The
simplest of such models is the MIS theory considered here (see also, e.g.,
Ref.~\cite{Luzum:2008cw}). More elaborate models could of course be considered
along the same lines, by identifying the transversely homogeneous attractor
locus and accounting for transverse dynamics at the linear level.

This paper is organized as follows. In \rfs{sec:background} we present the
variant of MIS theory which we adopt to illustrate the approach outlined above
and we describe the asymptotics of the background solutions at early as well as
late times.  Especially important are solutions on the attractor locus, which we
analyze in greater detail than can be found in the existing
literature. \rfs{sec:transverse} introduces perturbations around the attractor,
which bring in dependence on the transverse coordinates. The equations for
linearized perturbations are given and their late-time asymptotics are studied
and found to agree very well with numerical calculations already at moderate
times. One result of this asymptotic analysis is the precise form of the
suppression of short wavelength modes. We use this fact to implement a simple
numerical scheme whereby a finite set of modes is evolved and used to
reconstruct the spacetime evolution of an initial QGP drop.
In \rfs{sec:observables} we demonstrate that our approach qualitatively captures
effects of the transverse collective expansion, such as elliptic flow.
This is somewhat reminiscent of the effectiveness of the close limit in studies
of black hole collisions~\cite{Price:1994pm}. In \rfs{sec:outlook} we offer 
a summary of our findings along with
some remarks on how the approach presented here might be developed further.

\section{MIS theory and the attractor background}
\label{sec:background}

The MIS theory is expressed in terms of the classical
fields $\edens$ (the energy density), $u^\mu$ (the flow velocity) and $\pimunu$
(the shear-stress tensor). They satisfy the following set of partial
differential equations
\begin{subequations}\label{eq:MIS}
\begin{align}
  u\cdot\nabla\edens&=-(\edens+p)\nabla\cdot u+ u^\nu\nabla^\mu\pi_{\mu\nu},\\
  (\edens+p)u\cdot\nabla u_\mu&=-\Delta_{\mu\nu}\nabla^\nu p-\Delta_{\mu\nu}\nabla_\lambda\pi^{\nu\lambda},\\
 \Delta_{\mu\alpha}\Delta_{\nu\beta}u\cdot\nabla\pi^{\alpha\beta}&=-\left(1+\frac{4}{3}\tau_\pi\nabla\cdot u\right)\pi_{\mu\nu}-2\eta\sigma_{\mu\nu},
\end{align}
\end{subequations}
where $\nabla_\mu$ is the covariant derivative, $\Delta_{\mu\nu}\equiv
g_{\mu\nu}+u_\mu u_\nu$ is the transverse projector,
$\sigma_{\mu\nu}=\frac{1}{2}\Delta_{\mu\alpha}\Delta_{\nu\beta}(\nabla^\alpha
u^\beta+\nabla^\beta u^\alpha-\frac{2}{3}\Delta^{\alpha\beta}\nabla\cdot u)$ is
the shear tensor, $\eta$ is the shear viscosity and $\tau_\pi$ is the relaxation
time for $\pi_{\mu\nu}$. Throughout this paper we assume an equation of state
and transport coefficients dictated by conformal invariance:
\be
\label{eq:conformal}
\edens=\frac{1}{3}p=C_e T^4, \quad \eta=\frac{4}{3}C_e C_\eta T^3, \quad \tau_\pi=C_\tau T^{-1},
\ee
where $C_e$, $C_\eta$, and $C_{\tau}$ are dimensionless, constant transport
coefficients (see e.g.~\cite{Florkowski:2017olj}) and $T$ is the effective
temperature. For $\mathcal N=4$ supersymmetric Yang-Mills theory we have
$C_e=8\pi^2/15, C_\eta=1/4\pi, C_{\tau}=(2-\ln
2)/2\pi$~\cite{Bhattacharyya:2007vjd}. These values often serve as a point of
reference, and we have adopted them in our numerical calculations.

The basic physical picture we adopt is that of the Bjorken
model~\cite{Bjorken:1982qr}: at sufficiently high energies, in the first
approximation, the system exhibits boost invariance in the longitudinal
direction and homogeneity in the transverse directions (perpendicular to the
collision axis $z$). We will refer to this  approximate description as the
background; perturbations dependent on the transverse coordinates will
subsequently be treated at the linearized level. Under these assumptions, the
energy-momentum tensor of the conformal MIS theory can be parametrized in terms
of only two functions of proper time $\tau=\sqrt{t^2-z^2}$: the effective
temperature $T(\tau)$, and the pressure anisotropy $\pa(\tau)= 9
\pi^i_i/2\edens$, where $i=1,2$ labels the transverse coordinates (for details
see~\cite{Jankowski:2023fdz}). \rfes{eq:MIS} then reduce to
\begin{subequations}\label{eq:misbj}
\begin{align}
\tau\partial_\tau\ln T(\tau) &=-\f{1}{3}+\f{1}{18}\pa(\tau),\label{eq:misbj_T}\\
\tau\partial_\tau\pa(\tau) &=8\alpha^2 -\frac{\tau T(\tau)}{C_\tau} \pa(\tau)- \f{2}{9} \pa^2(\tau),\label{eq:misbj_A}
\end{align}
\end{subequations}
where
\begin{equation}
\alpha\equiv \sqrt{\frac{C_\eta}{C_{\tau}}}.
\end{equation}
In the NS limit ($C_\tau\to0$), \rfe{eq:misbj_A} becomes algebraic and one finds the well-known solutions
\begin{subequations}\label{eq:nsbj}
\begin{align}
T_{\rm NS}(\tau)&=\Lambda(\Lambda\tau)^{-\frac{1}{3}}\left(1-\frac{2C_\eta}{3}(\Lambda\tau)^{-\frac{2}{3}}\right),\label{eq:nsbjT} \\
\pa_{\rm NS}(\tau)&=\frac{8C_\eta}{\tau T(\tau)}\sim 8C_\eta(\Lambda\tau)^{-\frac{2}{3}}\left(1+\sum_{n=1}^\infty\left(\frac{2C_\eta}{3}\right)^n(\Lambda\tau)^{-\frac{2n}{3}}\right),\label{eq:nsbjA}
\end{align}
\end{subequations}
where $\Lambda$ is an integration constant with the dimension of energy.

General solutions of \rfes{eq:misbj} are not available in closed form and can
only be found in certain regimes. However, there are some special (and exact)
solutions which are completely independent of any integration constants:
$T=0,\pa=\pm 6\alpha$ and $T=\frac{2(4-\alpha^2)}{3C_\tau\tau},
\pa=-12$. We
regard these solutions as unphysical, since they require fine-tuning of the
initial conditions. To understand physically interesting solutions, we will
resort to approximations valid at small or large proper times.

At early times \rfes{eq:misbj} admits two special families of  solutions which can be characterized by having a finite value of the pressure anisotropy at $\tau=0$:
\begin{subequations}\label{eq:attractor}
\begin{align}
T_\pm(\tau)&\sim
\mu(\mu\tau)^{-\frac{1}{3}(1\mp\alpha)}\left(1\pm
\sum_{n=1}^{\infty} t_n^{(0)} (\mu\tau)^{\frac{n}{3}(2\pm\alpha)}\right), \\
\pa_\pm(\tau)&\sim\pm 6\alpha \left(1\pm
\sum_{n=1}^{\infty} a_n^{(0)} (\mu\tau)^{\frac{n}{3}(2\pm\alpha)}\right),
\end{align}
\end{subequations}
where $\mu$ is an integration constant. The early-time series  coefficients
$t_n^{(0)}$ and $a_n^{(0)}$ are rational functions of the transport coefficients
and the first few are given by
\begin{subequations}\label{eq:earlycoeff}
\begin{gather}
    t^{(0)}_1=-\frac{3\alpha}{C_{\tau}(2+\alpha)(2+9\alpha)}, \quad t^{(0)}_2=\frac{9\alpha(2+9\alpha+12\alpha^2)}{2C_{\tau}^2(2+\alpha)^2(2+5\alpha)(2+9\alpha)^2}, \quad \cdots\\
a^{(0)}_1=\frac{18\alpha}{C_{\tau}(2+9\alpha)},\quad a^{(0)}_2=\frac{54\alpha(2+7\alpha+7\alpha^2)}{C_{\tau}^2(2+\alpha)(2+5\alpha)(2+9\alpha)^2}, \quad \cdots
\end{gather}
\end{subequations}
The power series appearing in \rfes{eq:attractor} have a finite radius of convergence.
The upper sign in \rfes{eq:attractor} defines the class of attractor solutions labeled by $\mu$. It is easy to check that all these solutions
are mapped to the universal attractor  introduced in Ref.~\cite{Heller:2015dha} (see also the reviews~\cite{Florkowski:2017olj,Soloviev:2021lhs,Spalinski:2022cgj,Jankowski:2023fdz}).
The solutions with the lower sign are mapped to the ``repulsor'' solution noted in the original approach of Ref.~\cite{Heller:2015dha} (see also the recent Ref.~\cite{Aniceto:2022dnm}).

There are also ``generic'' solutions, characterized by a pressure anisotropy which diverges at early times. These solutions approach the attractor already in the far from equilibrium regime. To understand this, it is important to keep in mind that since \rfes{eq:misbj} are nonautonomous, the full phase space of solutions is three-dimensional and can be naturally parametrized by $(\tau,T,\tau T')$ \cite{Heller:2020anv} (see also Refs.~\cite{Spalinski:2022cgj,Jankowski:2023fdz}).
The attractor is a two-dimensional surface in this full phase space. In any constant proper-time slice of the phase space, $\mu$ is a parameter that labels points along the attractor curve on that slice. The generic solutions viewed on a sequence of such
phase-space slices at increasing values of $\tau$ gravitate toward this locus,
as a consequence of the fast longitudinal expansion at early times. This is illustrated in \rff{fig:bg}.

\begin{figure}[t]
\centering
\hspace{-0.02\textwidth}
\begin{minipage}{0.4\textwidth}
\includegraphics[width=1\textwidth]{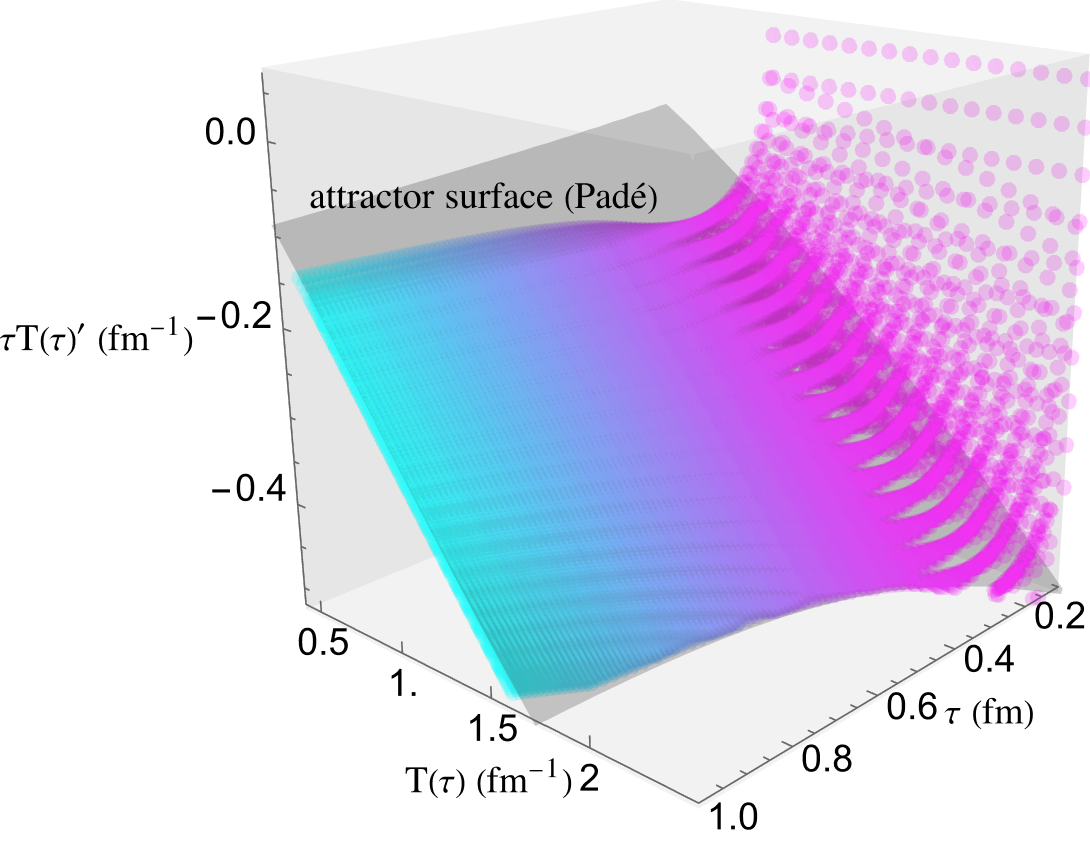}
\end{minipage}
\hspace{0.04\textwidth}
\begin{minipage}{0.55\textwidth} \includegraphics[width=1\textwidth]{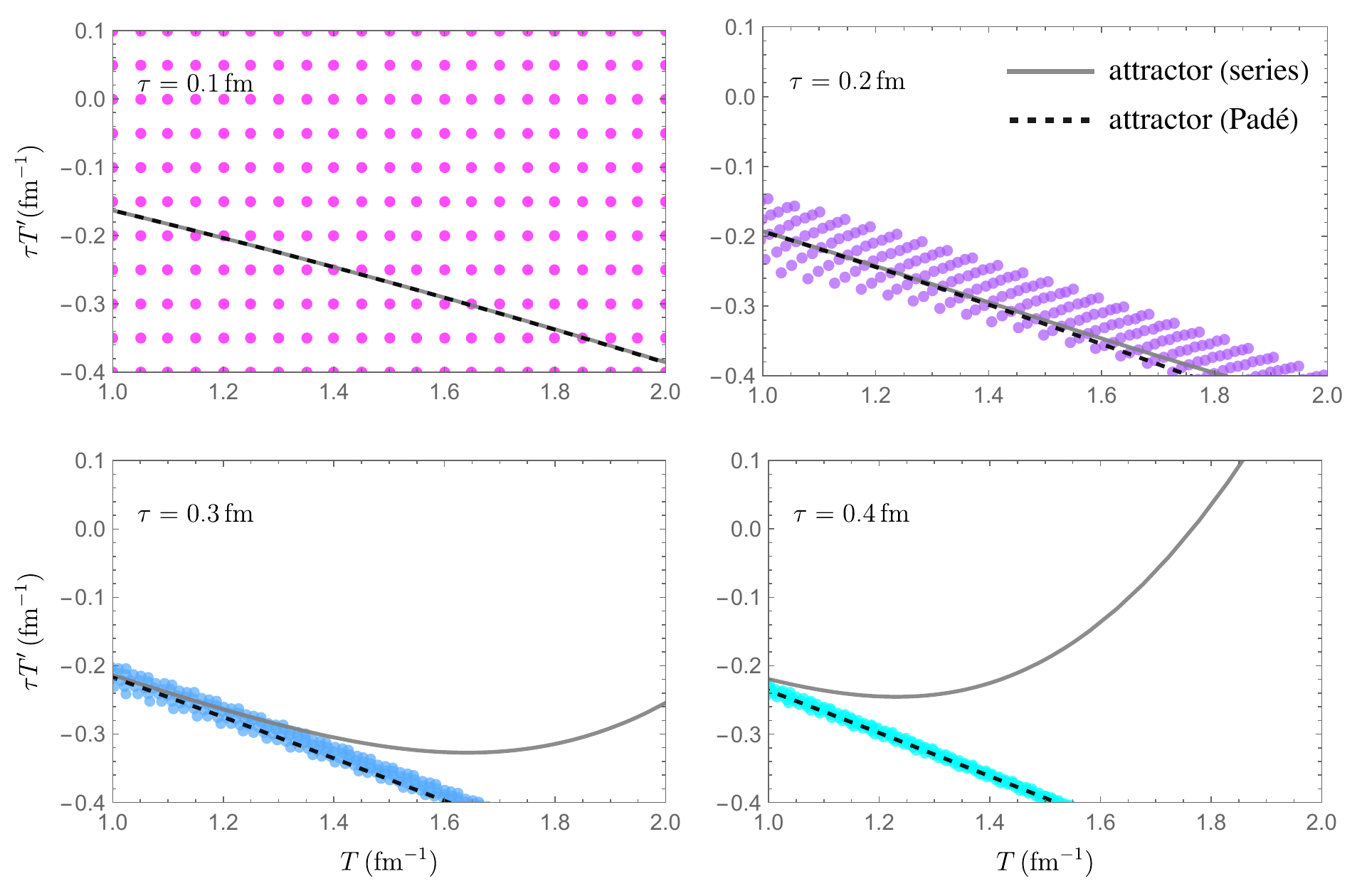}
\end{minipage}
\caption{The left panel shows the early-time attractor in the 3D phase space $(\tau, T, \tau T')$ with uniformly distributed initial conditions (points) at $\tau=0.1$ fm. The 2D attractor surface is obtained using Padé approximation of the series in \rfes{eq:attractor}. The right panel are the snapshots of the 3D plot at different times, where the solid and dashed line represents the series solution and its Padé approximant respectively.}
 \label{fig:bg}
\end{figure}

At late times, we expect the system to approach equilibrium in a way consistent with the original insights of Bjorken~\cite{Bjorken:1982qr}. In that regime, solutions to \rfes{eq:misbj}
can be represented in the form of transseries
\begin{subequations}\label{eq:bjorken}
\begin{align}
T(\tau) &\sim \Lambda\left[
(\Lambda\tau)^{-\frac{1}{3}}\left(1+\sum_{n=1}^\infty t_n^{(\infty)}(\Lambda\tau)^{-\frac{2n}{3}}\right)+C_{\infty}(\Lambda\tau)^{\frac{2}{3}\alpha^2-1}e^{-\frac{3}{2C_\tau}(\Lambda\tau)^{\frac{2}{3}}}\left(1+\mathcal O((\Lambda\tau)^{-\frac{2}{3}})\right) + \cdots \right] , \label{eq:bjorken_T}\\
\pa(\tau)&\sim 8C_\eta(\Lambda\tau)^{-\frac{2}{3}}\left(1+\sum_{n=1}^\infty a_n^{(\infty)}(\Lambda\tau)^{-\frac{2n}{3}}\right)+C_{\infty}'(\Lambda\tau)^{-\frac{2}{3}\alpha^2}e^{-\frac{3}{2C_\tau}(\Lambda\tau)^{\frac{2}{3}}}\left(1+\mathcal O((\Lambda\tau)^{-\frac{2}{3}})\right) + \cdots
\end{align}
\end{subequations}
These are of the form of an asymptotic power series augmented by an infinite set of exponential transseries contributions, of which
only the leading one is displayed above.
The first few coefficients of the late-time series $t_n^{(\infty)}$ and $a_n^{(\infty)}$ are given by
\begin{subequations}\label{eq:latecoeff}
\begin{gather}
    t^{(\infty)}_1=-\frac{2C_\tau\alpha^2}{3}, \quad
    t^{(\infty)}_2=-\frac{2C_\tau^2\alpha^2}{9},\quad \cdots \label{eq:latecoeff_t}\\
    a^{(\infty)}_1=\frac{2C_\tau}{3}(1+\alpha^2), \quad
    a^{(\infty)}_2=\frac{2C_\tau^2}{9}(4-\alpha^2+2 \alpha^4), \quad \cdots
\end{gather}
\end{subequations}
The power series appearing in \rfes{eq:bjorken} have a vanishing radius of convergence and are best interpreted
in the sense of asymptotic analysis -- through optimal truncation or by Borel summation (see, e.g. \rfcs{Aniceto:2015mto,Aniceto:2022dnm}).
    Note that in the limit $C_\tau\to0$ the leading and next-to-leading terms in
\rfes{eq:bjorken} reduce to their NS form,  given in \rfe{eq:nsbj},
upon substituting the coefficients in  \rfes{eq:latecoeff}.

    Initial conditions are mapped to the dimensionful scale $\Lambda$ and
    the dimensionless integration constant $C_\infty$ (the constant $C'_\infty$
    is not independent).
Given a specific solution, the corresponding values of $\Lambda$ and $C_\infty$ can be obtained
by fitting the leading terms in \rfe{eq:bjorken} at late-times.
For initial conditions relevant to heavy-ion physics one typically finds $\Lambda\approx\mathcal O(1)~\text{fm}^{-1}$.

The description of the attractor presented above differs from the original formulation of Ref.~\cite{Heller:2015dha} in that it uses proper time as the evolution parameter, rather than the dimensionless
evolution variable $\tau T$. This is natural when the dynamic system possesses scales in addition to temperature, as will be discussed in \rfs{sec:transverse}.
It is also worth emphasizing that while the attractor is often referred to as hydrodynamic, it coincides with hydrodynamics only at late times. At earlier, pre--hydrodynamic, times it depends on the nonhydrodynamic content
of the microscopic theory under consideration.

\section{The transverse perturbations}

In the previous section, we discussed the Bjorken background solution, which is an
idealized description where all the fields are homogeneous in the transverse
plane. While this idealization provides a useful first approximation for
modeling heavy-ion collisions at sufficiently high energies, it cannot account
for physical observables which depend on the structure of the plasma in the
transverse plane. In this section, we relax the transverse homogeneity condition by considering additional fields which aim to model the transverse dynamics.
These additional fields arise a perturbations of the fully nonlinear
hydrodynamic equations \rfn{eq:MIS} around the attractor background.

\label{sec:transverse}

\subsection{The linearized equations}
\label{sec:linearized}

We will look for solutions that can be approximated by the boost-invariant and
translation-invariant background solution discussed in \rfs{sec:background} and
a perturbation depending also on the transverse coordinates: 
\be
\label{eq:perts}
T(\tau, \bm x) = T(\tau) + \delta T(\tau, \bm x), \quad
\umu(\tau, \bm x) =u^\mu+\delta\umu(\tau, \bm x), \quad
\pimunu(\tau, \bm x) = \pi^{\mu\nu}(\tau) + \delta\pimunu(\tau, \bm x),
\ee
where $\bm x=(x_1,x_2)$ labels the coordinates of the transverse plane. We shall
always retain the argument $\bm x$ of the above quantities to distinguish the
full ones from the background, which depends only on $\tau$ (and, as the only
argument, is often suppressed). The background fields $T$ and $\pi^{\mu\nu}$ are
taken to be on the attractor locus defined in \rfs{sec:background}, and
$u^\mu=(-1,\bm 0,0)$.  Due to the assumed symmetries and the transversality
condition $u^\mu\pi_{\mu\nu}=0$, the shear-stress tensor has only one
independent background component (i.e., $\pi^{11}=\pi^{22}$) and three
independent perturbation components (i.e., $\delta\pi^{11},\delta\pi^{22},
\delta\pi^{12}$) that couple to the perturbation of hydrodynamic fields $\delta
T$ and $\delta\bm u$.  We also put $\delta \pi^{i\eta}=0$, which is consistent
due to $\delta u^\eta=0$.

The perturbation fields can also be normalized by the background energy scale $T(\tau)$, i.e.,
\be
\label{eq:dimless1}
\delta\hat{T}(\tau,\bm x) = \frac{\delta T(\tau,\bm x)}{T(\tau)}, \quad \delta\hat{\pi}_{ij}(\tau,\bm x)=\frac{\delta\pi_{ij}(\tau,\bm x)}{C_eT(\tau)^4},
\ee
such that all six perturbation fields, collectively denoted by $\hat\phi(\tau, \bm x)=(\delta\hat T, \delta u_1, \delta u_2, \delta\hat\pi_{12}, \delta\hat\pi_{11}, \delta\hat\pi_{22})$, are dimensionless.
Since the background is independent of the transverse coordinates, it is also natural and convenient to introduce the Fourier transforms just for the perturbations:
\be
\label{eq:ftrafo}
\hat\phi(\tau, \bm x) = \int \frac{d^2 k}{(2\pi)^2}\ e^{\mathrm i \bm k\cdot\bm x}{\hat\phi}(\tau,\bm k),
\ee
where we retain the argument $\bm x$ or $\bm k$ to distinguish perturbation fields $\hat\phi$ in different Fourier spaces.

Linearization of the full MIS equations \rfn{eq:MIS} around an attractor
solution leads to a system of six linear partial differential equations for the
perturbations.  For each value of $\bm k$ the set of six modes
$\hat{\phi}(\tau,\bm k) \equiv (\delta \hat{T}, \delta u_1, \delta u_2, \delta
\hat{\pi}_{12}, \delta\hat{\pi}_{11}, \delta\hat{\pi}_{22})$ satisfies a linear
system of evolution equations which can be explicitly written as
\begin{subequations}\label{eq:linsys1}
\begin{align}
4\left(\tau\partial_\tau+\frac{2\pa}{9}\right)\delta\hat T+\frac{(12+\pa)}{9} \mathrm{i}\tau k_i\delta u^i-\delta\hat\pi^i_i&=0,\\
\frac{4}{3}\mathrm{i}\tau k_i\delta\hat T+\frac{1}{9}\left[(12+\pa)\tau\partial_\tau-4+8\alpha^2+\left(\frac{7}{3}-w\right)\pa\right]\delta u_i+\mathrm{i}\tau k^j\delta\hat\pi_{ij}&=0,\\[5pt]
\frac{1}{9}\left(w\pa-32\alpha^2\right)\delta_{ij}\delta\hat T+\frac{4}{27}\mathrm{i}\tau\left[\left(\pa-6\alpha^2\right)\delta_{ij}k_\ell+9\alpha^2(\delta_{i\ell}k_j+\delta_{j\ell}k_i)\right]\delta u^\ell&\nonumber\\
+\left(\tau\partial_\tau+w+\frac{2\pa}{9}\right)\delta\hat\pi_{ij}&=0,
\end{align}
\end{subequations}
where $w(\tau)=\tau T(\tau)/C_\tau$. Since modes with different $\bm k$ are decoupled at the linearized level,
we omit the arguments $\tau$ and $\bm k$ unless needed.

\subsection{Numerical evolution of the modes}
\label{sec:numermodes}

It is straightforward to numerically solve the system of ODEs given in
\rfes{eq:linsys1} together with \rfes{eq:misbj} for any given wave vector $\bm
k$. To do this we need to choose a background solution $T(\tau), \pa(\tau)$ and
initial conditions for the modes. In practice, one would take an initial
condition for all the hydrodynamic fields taken from some model of the initial
state and represent them as a homogeneous background plus a perturbation as in
\rfes{eq:linsys1}. The initial values for the modes can then be obtained from this by
calculating the Fourier transform of the spacetime perturbations with respect to
the transverse coordinates. The Fourier
modes can then be evolved by solving the system \rfes{eq:linsys1} for each mode.
To reconstruct the spacetime fields one then needs to invert the Fourier
transform. In practice this has to be done numerically by discretizing the
transverse space (and the corresponding space of wave vectors). We will discuss
how this can be done in more detail in \rfs{sec:observables}, but in this
section we will only consider how the dynamics of the Fourier modes
depends on the wave vectors~${\bm k}$.

The results of such a numerical calculation are shown in \rff{fig:damp}.
The most obvious feature of the results is the fact that modes with
higher $k$ are damped more strongly than modes with low $k$, as  illustrated in
the left panel of \rff{fig:damp}. The origin of this phenomenon can be
understood analytically, as will be discussed in \rfs{sec:asymptotic}. Moreover,
one can also see that
the perturbations starting away from the attractor are damped more significantly
than perturbations starting on the attractor,  as  illustrated in the
right panel of \rff{fig:damp}. This suggests the stability of the
linearization around the attractor background
and reveals its dominant role in the presence of transverse dynamics.
We will refine this statement based on
analytic results presented in \rfs{sec:asymptotic}.

\begin{figure}[t!]
\centering
\hspace{-0.02\textwidth}
\begin{minipage}{0.48\textwidth} \includegraphics[width=1\textwidth]{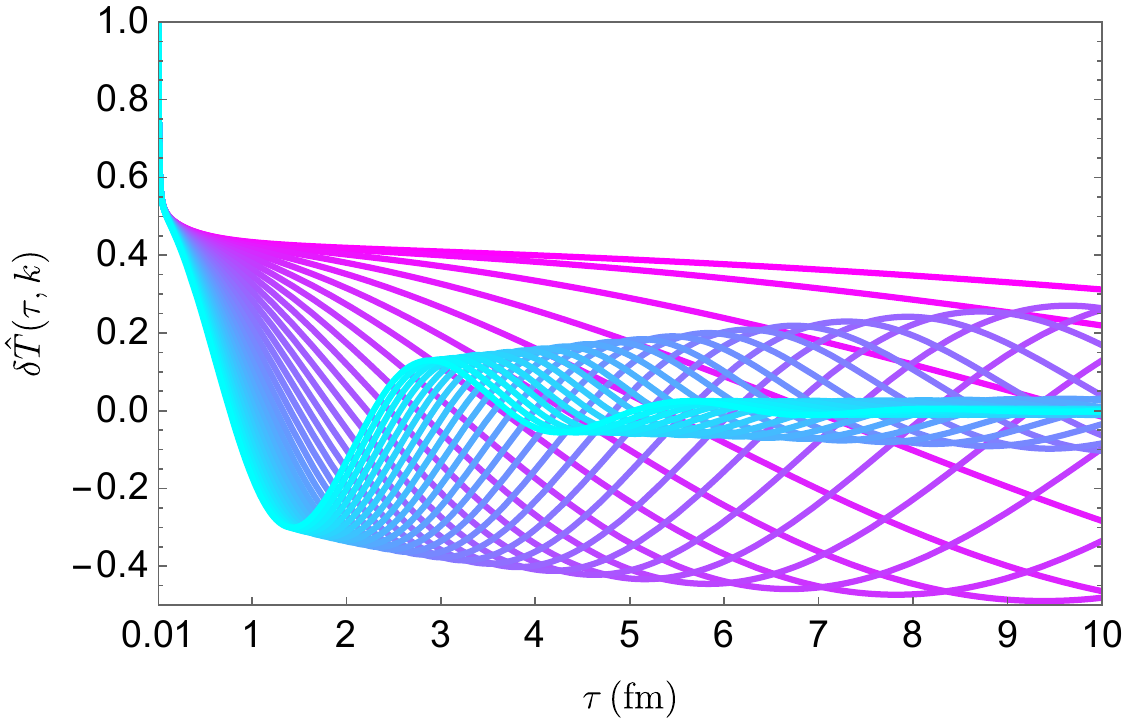}
\end{minipage}
\hspace{0.03\textwidth}
\begin{minipage}{0.48\textwidth} \includegraphics[width=1\textwidth]{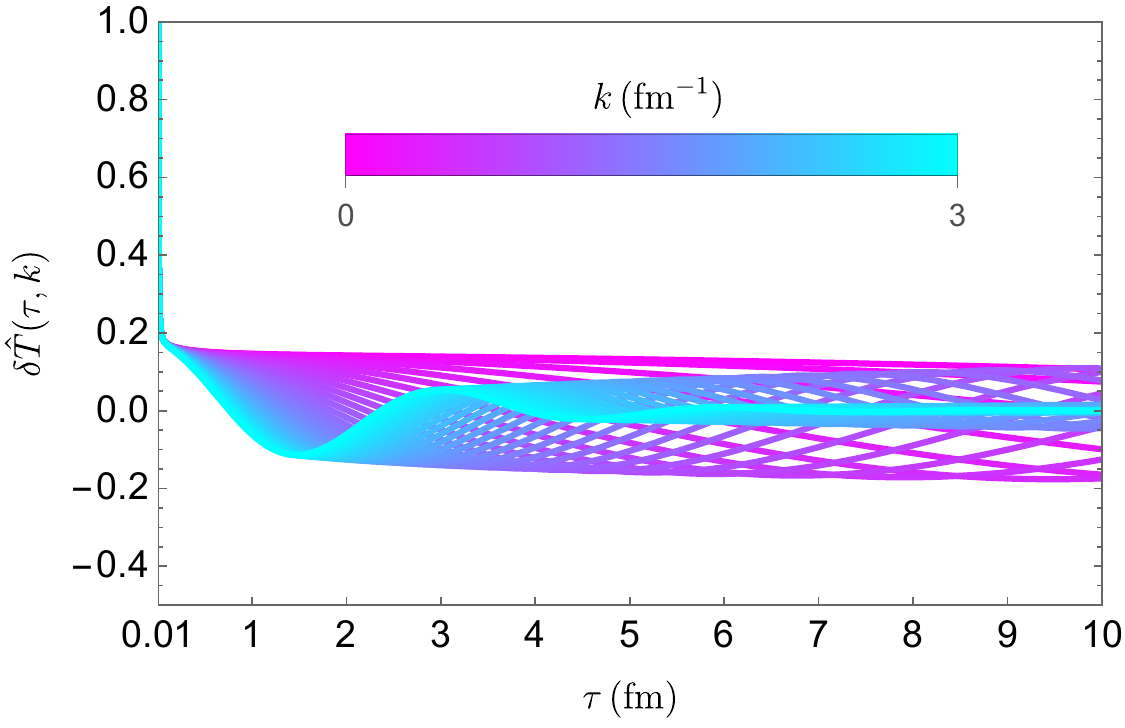}
\end{minipage}
\caption{Evolution of Fourier modes with various values of $k$ between $0 - 3\,
    \text{fm}^{-1}$, with the background initially resided on the attractor,
    $\pa(0)\simeq6\alpha$ (left) and off it, $\pa(0)\simeq30\alpha$ (right), with initial values of $\delta\hat T$ fixed to 1 for different values of $k$. }
 \label{fig:damp}
\end{figure}

\subsection{Late-time behavior of perturbations }
\label{sec:asymptotic}

The main goal of this section is to understand analytically the main
features of the numerical solutions of the mode equations \rfes{eq:linsys1}
described in \rfs{sec:linearized}, especially the stability of the Bjorken
attractor background. To this end, we will study  the late proper-time behavior
of the modes $\hat{\phi}(\tau,\bm k)$. The set of six ODEs describing the
evolution of the modes $\hat{\phi}$ can be analyzed by standard asymptotic
methods, although the complexity of this problem makes the calculation
technically nontrivial. In doing this it is important to recognize that the case
when ${\bm k}=0$ must be treated separately, do to the presence of products such
as ${\bm k}\tau$ in the mode evolution equations. We will first consider the
case when ${\bm k}\neq 0$, returning to the issue of the zero-modes at the end
of this section.

For the purpose of calculating the asymptotics it is convenient to introduce the transverse divergence $\delta\theta(\tau,\bm x)\equiv\partial_i\delta u^i$ and longitudinal vorticity $\delta\omega(\tau,\bm x)\equiv\epsilon_{ij}\partial^i\delta u^j$ where $\epsilon_{ij}$ is the Levi-Civita symbol.
In $\bm k$ space we will use the following dimensionless quantities, normalized by $k\equiv |\bm k|$:
\be
\label{eq:dimless2}
\delta\hat\theta(\tau,\bm k)=\frac{\delta{\theta}(\tau,\bm k)}{k}=\mathrm{i}\hat k_i\delta u^i(\tau, \bm k), \quad \delta\hat\omega(\tau,\bm k)=\frac{\delta{\omega}(\tau,\bm k)}{k}=\mathrm{i}\epsilon_{ij}\hat k^i\delta u^j(\tau, \bm k),
\ee
where $\hat k_i\equiv k_i/k$. For such a choice of variables,
the linear system of evolution equations reads
\begin{subequations}\label{eq:linsys2}
\begin{align}
    4\left(\tau\partial_\tau+\frac{2\pa}{9}\right)\delta\hat T+\frac{(12+\pa)}{9}\tau k\delta\hat\theta-\delta\hat\pi^i_i&=0,\label{eq:linsys2-1}\\
    -\frac{4}{3}k^2\delta\hat T+\frac{1}{9\tau}\left[(12+\pa)\tau\partial_\tau-4+8\alpha^2+\left(\frac{7}{3}-w\right)\pa\right]k\delta\hat\theta-k_ik_j\delta\hat\pi^{ij}&=0,\label{eq:linsys2-2}\\
    \frac{1}{9\tau}\left[(12+\pa)\tau\partial_\tau-4+8\alpha^2+\left(\frac{7}{3}-w\right)\pa\right]k\delta\hat\omega-\epsilon_{ij}k^jk^\ell\delta\hat\pi^i_\ell&=0,\label{eq:linsys2-3}\\
    \frac{1}{9}\left(w\pa-32\alpha^2\right)\delta_{ij}\delta\hat T+\frac{4\tau}{27k}\left[18\alpha^2k_ik_j+\left(\pa-6\alpha^2\right)k^2\delta_{ij}\right]\delta\hat\theta&\nonumber\\
    -\frac{4\alpha^2(k_i\epsilon_{j\ell}+k_j\epsilon_{i\ell})k^\ell\tau}{3 k}\delta\hat\omega+\left(\tau\partial_\tau+w+\frac{2\pa}{9}\right)\delta\hat\pi_{ij}&=0.\label{eq:linsys2-4}
\end{align}
\end{subequations}
In these equations the quantities $T$ and $\pa$ refer to the background solutions, expanded as in \rfe{eq:bjorken}. As discussed earlier, this late-time asymptotics of the background depends on the scale $\Lambda$ which is determined by the initial conditions of the background. This scale will therefore appear also in the asymptotics of the perturbations.

At this juncture, one way to proceed is to rewrite this system in terms of
higher-order ODEs;  remarkably, it can be written as a set of three second-order
ODEs for $\delta\hat T, \delta\hat\theta, \delta\hat\omega$ -- they are given in
\rfa{app:ODE}. Two of these equations couple $\delta\hat T$ and
$\delta\hat\theta$, while the third involves $\delta\hat\omega$ alone. The two
coupled equations can be combined into a fourth-order ODE for $\delta\hat T$,
whose asymptotic behavior can be studied by standard methods (see, e.g.,
Ref~\cite{Bender78:AMM}). This is in principle straightforward, but is somewhat
challenging in practice due to the complexity of the coefficients which appear
in this equation. Alternatively, one can analyze the system of six first-order
ODEs directly using the methods developed in Ref.~\cite{wasow1965asymptotic}.
Both approaches gives rise to the same late-time asymptotic solutions whose {\em
leading} terms take the form:
\begin{subequations}
\label{eq:MIS_finitek}
\begin{align}
    \delta\hat T& \sim \sum_{i=1}^4C_i(\Lambda\tau)^{\beta_i}e^{-\mathrm{i}\omega_i \tau-A_i(\Lambda\tau)^{\frac{2}{3}}}\left(1+\mathcal O((\Lambda\tau)^{-\frac{2}{3}})\right),
    \\
    \delta\hat \theta &\sim\sum_{i=1}^4C'_i(\Lambda\tau)^{\beta'_i}e^{-\mathrm{i}\omega_i\tau-A_i(\Lambda\tau)^{\frac{2}{3}}}\left(1+\mathcal O((\Lambda\tau)^{-\frac{2}{3}})\right),
    \\
    \delta\hat \omega &\sim\sum_{i=5}^6C_i(\Lambda\tau)^{\beta_i}e^{-\mathrm{i}\omega_i\tau-A_i(\Lambda\tau)^{\frac{2}{3}}}\left(1+\mathcal O((\Lambda\tau)^{-\frac{2}{3}})\right).
\end{align}
\end{subequations}
The initial conditions are accounted for by the amplitudes $C_1,\dots,C_6$ and $C_1',\dots,C_4'$. The primed integration constants are related to the unprimed ones by the relations
\begin{align}\label{eq:CC'}
C_1'=3 \mathrm{i} \cmis C_1, \quad
C_2' =-3 \mathrm{i} \cmis C_2, \quad
C_3'=-\frac{(1-3\alpha^2)\Lambda}{2\alpha^2k}C_3, \quad
C_4' =\frac{\Lambda}{\cmis^2 C_{\tau} k}C_4,
\end{align}
so that only the unprimed integration constants $C_1,\dots,C_6$ are independent. Recall also that each of the perturbations appearing above depends on the wave vector $\bm k$, and so do the coefficients $C_1,\dots,C_6$.

\begin{figure}
\centering
\includegraphics[width=0.62\textwidth]{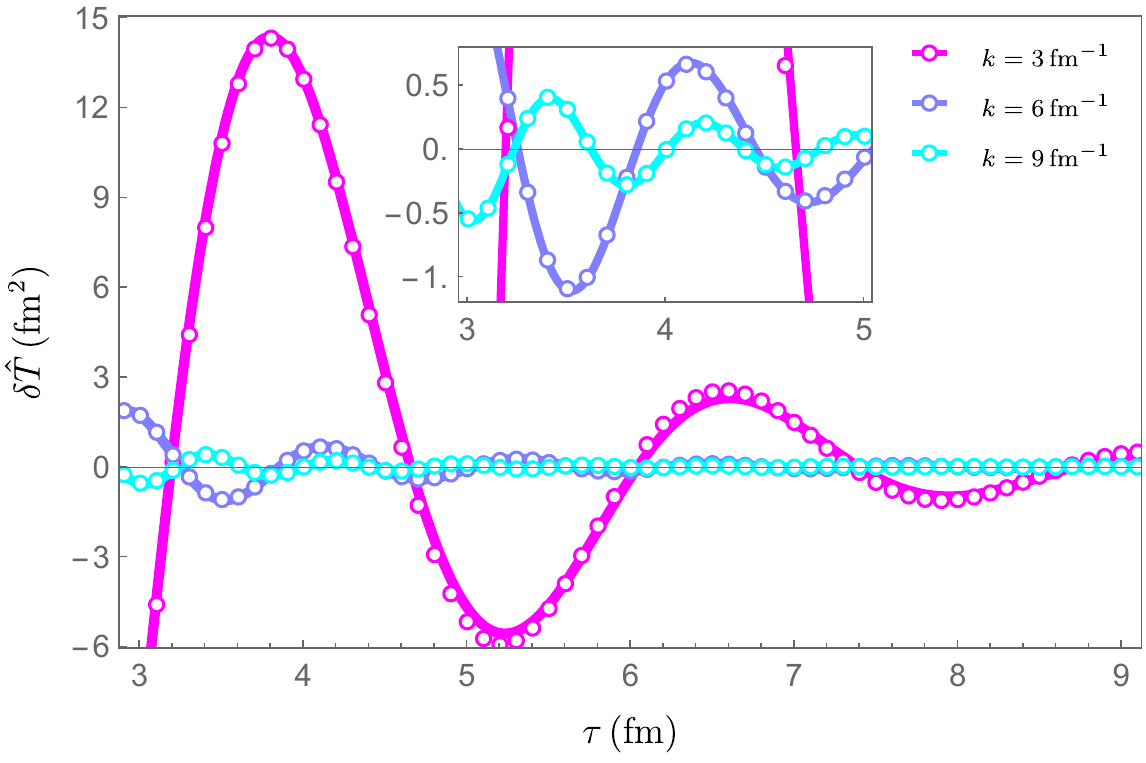}
\\
\vspace{5pt}
\includegraphics[width=0.62\textwidth]{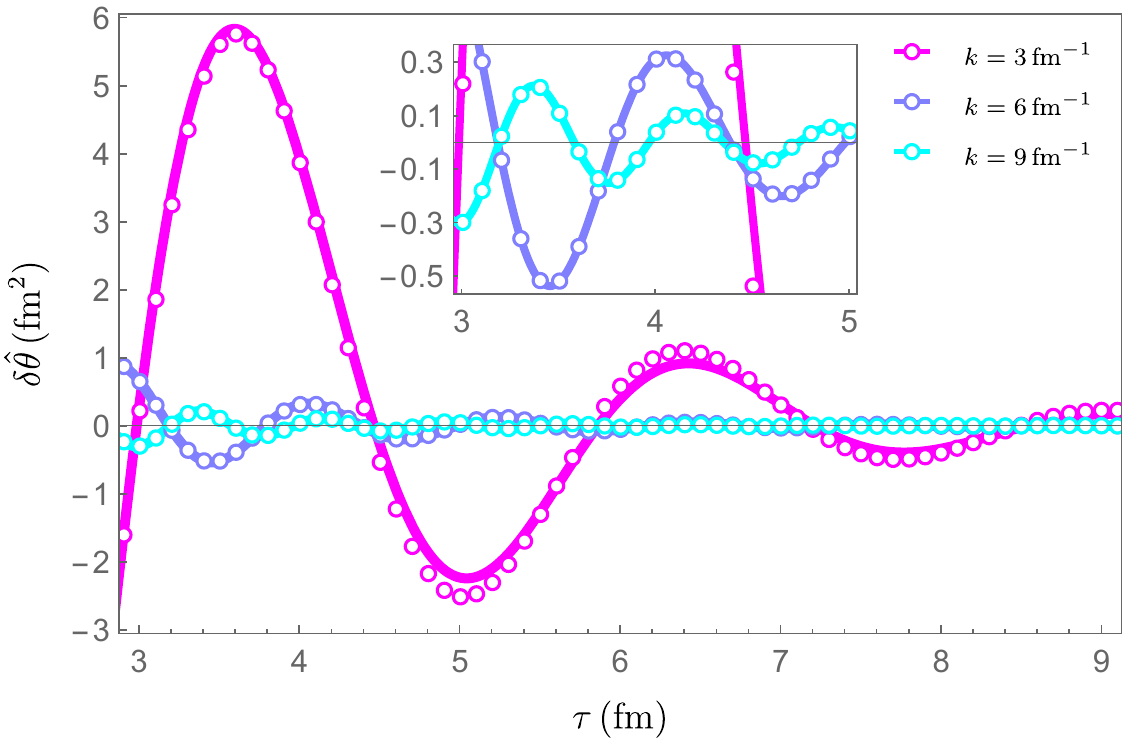}
\\
\vspace{5pt}
\hspace{-10pt}
\includegraphics[width=0.631\textwidth]{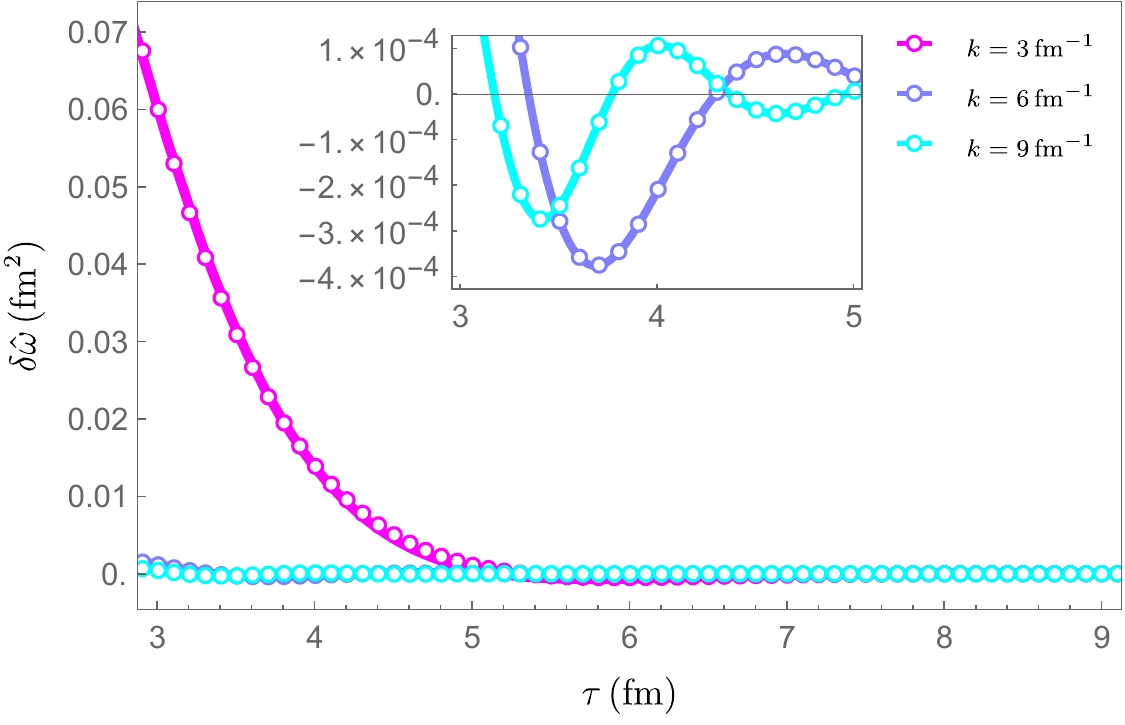}
\caption{The figures show the analytic late-time asymptotic solutions (solid lines) and the numerical results obtained in \rfs{sec:linearized} (discrete points) for various values of wave vector $k$. The inset plots are provided to give a better view of the small-$\tau$ region. The agreement is better for higher values of $k$; at low values one could nevertheless improve it by incorporating  subleading terms in \rfe{eq:MIS_finitek}.
}
 \label{fig:asymptotics}
\end{figure}

The remaining quantities appearing in \rfes{eq:MIS_finitek} above are determined in terms of the parameters of the theory and do not depend on the initial state. The parameters appearing in the exponentials are
given by
\begin{gather}
A_1=A_2= \frac{\alpha^2}{ C_\tau \cmis^2}, \quad A_3=\frac{3}{2C_\tau}, \quad A_4=\frac{1}{2C_\tau \cmis^2}, \quad
A_5=A_6= \frac{3}{4C_\tau},\nonumber\\
\omega_1=-\omega_2=\cmis k\left[1+\frac{2\alpha^2}{3\cmis^2}\left(2C_{\tau}(1-\alpha^2)-\frac{(1+\alpha^2)\Lambda^2}{C_{\tau}^2\cmis^4k^2}\right)(\Lambda\tau)^{-\frac{2}{3}}\right], \quad \omega_3=\omega_4=0,\nonumber\\
\omega_5=-\omega_6=\alpha k\left[1-\left(C_{\tau}\alpha^2+\frac{3\Lambda^2}{8C_{\tau}^2\alpha^2k^2}\right)(\Lambda\tau)^{-\frac{2}{3}}\right],
\end{gather}
where
\be
\cmis
=\sqrt{\f{1}{3}(1+4\alpha^2)}
\ee
is the asymptotic speed of sound of MIS theory, which remains subluminal as long
as $\alpha^2<1/2$ (or $C_\tau > 2 C_\eta$). The constant coefficients $A_i$ are the leading
contributions to eigenvalues of the coupled linear equations describing
perturbations around the attractor. They are real, positive and independent of
the wave vectors $k$, implying stability of the attractor against transverse
perturbations regardless of their amplitudes. Note that the quantities
$\omega_1=-\omega_2$ and $\omega_5=-\omega_6$ are actually functions of $\tau$,
which also have a nontrivial dependence on the wave vectors $k$. At late times
these quantities approach frequencies of oscillation which become harmonic in
that domain.

The coefficients appearing in the power-law factors in \rfes{eq:MIS_finitek} are given by
\begin{gather}
\beta_1=\beta_2=\beta'_1=\beta'_2=\frac{1}{54\cmis^4}\left(1+8\alpha^2+64\alpha^4+32\alpha^6+\frac{4\alpha^2\Lambda^2}{C_{\tau}^3\cmis^4k^2}\right), \quad \beta_3=\beta'_3+1 =-\frac{2}{3}(1-\alpha^2),  \nonumber\\
\beta_4=\beta'_4+\frac{1}{3}=\frac{2\alpha^2}{27\cmis^4}\left(1-16\alpha^2-\frac{2\Lambda^2}{C_{\tau}^3\cmis^4k^2}\right), \quad \beta_5=\beta_6=\frac{1}{6}(1+2\alpha^2),\label{eq:coeff}
\end{gather}
A key point which follows from these relations is the suppression of modes with
large $k$ relative to modes with smaller values of $k$. This fact was already
noted in the numerical solutions discussed in
  \rfs{sec:numermodes} (see also \rff{fig:damp}), but the asymptotic formulas \rfes{eq:MIS_finitek} in conjunction with \rfe{eq:coeff} reveal the precise form
of this suppression. Note also that the solution labeled by $i=3$ reproduces
the transseries solution describing the homogeneous background, given by
\rfes{eq:bjorken} and \rfn{eq:MIS_zerok} (with the integration constant
independent of $k$ in that case).

This highly nontrivial structure very accurately reproduces the behavior seen in
the numerical calculations described in \rfs{sec:numermodes}. At late time, the
dominant asymptotic solutions for $\delta\hat T$ and $\delta\hat\theta$ are
those which are least damped, i.e. those with the smallest value of $A_i$. For
the $\mathcal N=4$ supersymmetric Yang-Mills theory these are terms labeled by
$i=1,2$ in the solutions of \rfes{eq:MIS_finitek}. These formulas then capture only
the leading exponential behavior, but still give an excellent account of the
numerical solutions, as illustrated in \rff{fig:asymptotics}, especially for
large and moderate values of the wave vector $k$. These plots were obtained by
matching the complex amplitude $C_1$ to initial data; the amplitude $C_2$ is
determined by conjugation, since the solutions are real.    It is also quite
interesting to compare these results with the asymptotics of Navier-Stokes
theory, as well as with the case of ideal fluids. This is discussed in
\rfa{app:NS-ID}.

The asymptotic solution discussed above applies to modes with $k\neq0$. We now
turn to the $k=0$ modes, which describe perturbations homogeneous in the
transverse space. It is more convenient to study this case using
\rfes{eq:linsys1} which reduce to
\begin{subequations}\label{eq:zerok}
\begin{align}
    \left(\partial_\tau+\frac{2\pa}{9\tau}\right)\delta\hat T-\frac{1}{4\tau}\delta\hat\pi_i^i&=0,\\
    \left[(12+\pa)\tau\partial_\tau-4+\frac{8C_\eta}{C_\tau}+\left(\frac{7}{3}-\frac{\tau T}{C_\tau}\right)\pa\right]\delta u_i&=0,\\
    \frac{1}{9}\left(\tau T\pa-32C_\eta\right)\delta_{ij}\delta\hat T+\left(C_\tau\tau\partial_\tau+\tau T+\frac{2C_\tau\pa}{9}\right)\delta\hat\pi_{ij}&=0.
\end{align}
\end{subequations}
We observe that $\delta u_i$ and $\delta
\hat\pi_{12}$ decouple from all other modes, and the only remaining coupling is between  $\delta\hat T$ and $\delta\hat\pi_{ij}$. The solutions in the late-time expansion are given by
\begin{align}\label{eq:MIS_zerok}
\delta u_i(\tau)&\sim  C_i
(\Lambda\tau)^{\frac{1}{3}}\left(1+2\alpha^2C_\tau(\Lambda\tau)^{-\frac{2}{3}}+\frac{4(1+6\alpha^2)\alpha^2C_\tau^2}{9}(\Lambda\tau)^{-\frac{4}{3}}+\mathcal O((\Lambda\tau)^{-2})\right),\nonumber\\
    \delta{\hat T}(\tau)&\sim C_3 \left(1+\frac{2\alpha^2C_\tau}{3}(\Lambda\tau)^{-\frac{2}{3}}+\frac{4\alpha^2(1+\alpha^2)C_\tau^2}{9}(\Lambda\tau)^{-\frac{4}{3}}+\mathcal O((\Lambda\tau)^{-2})\right)\nonumber\\
    &\quad+ C_4
    (\Lambda\tau)^{-\frac{2}{3}\left(1-\alpha^2\right)}e^{-\frac{3}{2C_\tau}\tau^{\frac{2}{3}}}\left(1+\mathcal O((\Lambda\tau)^{-\frac{2}{3}})\right),\nonumber\\
    \delta\hat\pi_{ii}(\tau)&\sim C_3\frac{8\alpha^2C_\tau}{3}(\Lambda\tau)^{-\frac{2}{3}}\left(1+\frac{4(1+3\alpha^2)C_\tau}{9}(\Lambda\tau)^{-\frac{2}{3}}+\mathcal O((\Lambda\tau)^{-\frac{4}{3}}\right)\nonumber\\
    &\quad+ \left(-C_4\frac{2}{C_\tau}+C_5(-1)^i\right)(\Lambda\tau)^{\frac{2}{3}\alpha^2}e^{-\frac{3}{2C_\tau}(\Lambda\tau)^{\frac{2}{3}}}\left(1+\mathcal O((\Lambda\tau)^{-\frac{2}{3}})\right),\nonumber\\
\delta\hat\pi_{12}(\tau)&\sim C_6(\Lambda\tau)^{\frac{2}{3}\alpha^2}e^{-\frac{3}{2C_\tau}(\Lambda\tau)^{\frac{2}{3}}}\left(1+\mathcal O((\Lambda\tau)^{-\frac{2}{3}})\right),
\end{align}
where $i=1,2$. Note that $\{C_n\}$ with $n=1,\dots 6$ is another set of
integration constants, distinct from those in \rfes{eq:MIS_finitek}. The second
solution in $\delta\hat T$ is nothing but the transseries solution of $T(\tau)$
in \rfes{eq:bjorken} (and also the solution of $\delta\hat T$ labeled by $i=3$
in \rfes{eq:MIS_finitek} if the integration constant is independent of $k$). The
above solutions also reproduce the transseries solution of $\pa(\tau)$ in
\rfes{eq:bjorken}, using $\delta\pa=(9\delta\hat\pi^i_i-8\pa\delta\hat T)/2$.

To summarize, at late times the characteristics of the initial state are mapped
to the scale $\Lambda$ and dimensionless amplitudes $\{C_n(k)\}$ which carry
information about the structure of the initial data in the transverse plane.
These numbers can be matched to a given numerical solution of \rfes{eq:linsys2}
and \rfn{eq:zerok}. All physical observables that depend on the transverse
dynamics can be expressed in terms of this asymptotic data, in manner described
explicitly in the following section.

We now turn to the stability of the  attractor in the presence of transverse perturbations. As seen in \rfes{eq:MIS_finitek} all nonzero modes are damped to zero at late times.
The same is true for the zero-modes, with the exception of the velocity
perturbations. As seen in \rfe{eq:MIS_zerok}, these grow as $\delta u_i(\tau,
{\bm k}|)_{{\bm k}=0} \sim\tau^{1/3}$ at late times. This will invalidate the
linear approximation at sufficiently late times, unless the initial condition
for this mode is zero -- in this special  case it remains zero in the course of
evolution, since (in contrast to nonzero modes) it is not sourced by other
modes. If the initial amplitude is nonzero but small enough, the mild growth
will not invalidate the linear approximation until after freeze-out. For
applications to heavy-ion collision it is usually assumed that all modes of the
velocity perturbation are negligible initially, which makes the picture
described here workable.

It is interesting to note that the $\tau^{1/3}$ behavior of the velocity
zero-mode originates in the perfect-fluid theory, where is it easily seen to
follow from energy-momentum conservation. It is likely that this is a general
property of perturbations of Bjorken flow regardless of the dynamics. Further
details concerning the perfect-fluid and NS limits are presented in
\rfa{app:NS-ID}.

\subsection{Spacetime evolution of QGP}
\label{sec:real}

We now turn to the problem of reconstructing the spacetime configuration at a
given time from the evolved Fourier modes. The first step is the choice of
background. This one wants to do in such a way as to be able to accommodate any
initial condition relevant to heavy-ion physics. In this work we assume that the
initial velocity perturbations vanish, along with the initial
$\delta\pi^{\mu\nu}$. What remains is the initial condition for the energy
density (or effective temperature) profile in the transverse plane.

Given an initial effective temperature profile $T(\tau_i, \bm x)$, the first task is to determine a suitable Bjorken background, that is, the initial temperature $T(\tau_i)$. The initial condition
is then split into two parts as required by \rfe{eq:perts}: the homogeneous background and a perturbation
\be
 T(\tau_i, \bm x) = T(\tau_i) + \delta T(\tau_i, \bm x) = T(\tau_i)(1 +
\delta \hat{T}(\tau_i, \bm x)),
\ee
Since we linearize in the perturbation, the validity of the above separation formally requires
\be
\label{lessone}
|\delta T(\tau_i,\bm x)|\lesssim T(\tau_i) \quad\text{or}\quad  |\delta\hat T(\tau_i,\bm x)|\lesssim 1.
\ee
One can ensure that this is satisfied at the initial time by  choosing the background temperature
\be
T(\tau_i) = T_\mathrm{max}/2
\ee
where $T_\mathrm{max}$ is the maximum temperature at the initial time. With this
choice, $\delta\hat T(\tau_i,\bm x)$ varies between $-1$ and $1$ when
$T(\tau_i,\bm x)$ varies  between $0$ and $T_\mathrm{max}$ in the transverse.
Thus, given an initial temperature profile in the transverse plane, we can always pick
the background such that our perturbation satisfies \rfe{lessone} at the initial
time. This condition persists at later times due to the damped
nature of the solutions. Thus our approach at least formally makes sense for any
initial effective temperature profile.

Once the background is chosen, the Fourier transform of $\delta\hat T(\tau_i,\bm
x)$ then provides initial conditions for the mode equations. Realistic initial
conditions are in principle superpositions of infinitely many Fourier modes, but
due to the strong damping of modes with high $k$, one can approximate the
initial state by keeping only some finite number of low-$k$ modes. In practice,
one needs to evaluate the initial conditions on a finite grid in the transverse
plane, calculate the Fourier transform, evolve the modes by solving
\rfes{eq:linsys1} and then compute the inverse Fourier transform to reconstruct
the spacetime history of the QGP. The problem can thus be described by a finite
set of modes which encode the transverse structure of the initial state. The
procedure can be made highly efficient by applying the Fast Fourier Transform
(see e.g. Ref.~\cite{PresTeukVettFlan92}).

In our pilot study we consider a computational domain which is a $40$ fm by $40$
fm square region in the transverse plane, described by a regular grid of $100$
by $100$ points. A corresponding (conjugate) grid is then constructed in $\bm
k$-space -- this also introduces a cutoff on large $k$. We have implemented the
steps outlined above for the case of very simple initial conditions which assume
$\delta T$ in the form of a Gaussian distribution in transverse coordinates with
a specified impact parameter $b$. The remaining fields are taken to vanish at
the initial time, which is a natural choice at least in the case of the
transverse velocity perturbation.
This leads to results shown in \rff{fig:gaussian}; they are analysed in more
detail in \rfs{sec:observables}.
One can of course apply this
approach to initial states generated in other ways, such as events generated
using  Glauber Monte-Carlo.

\begin{figure}[t]
\centering
\includegraphics[width=0.98\textwidth]{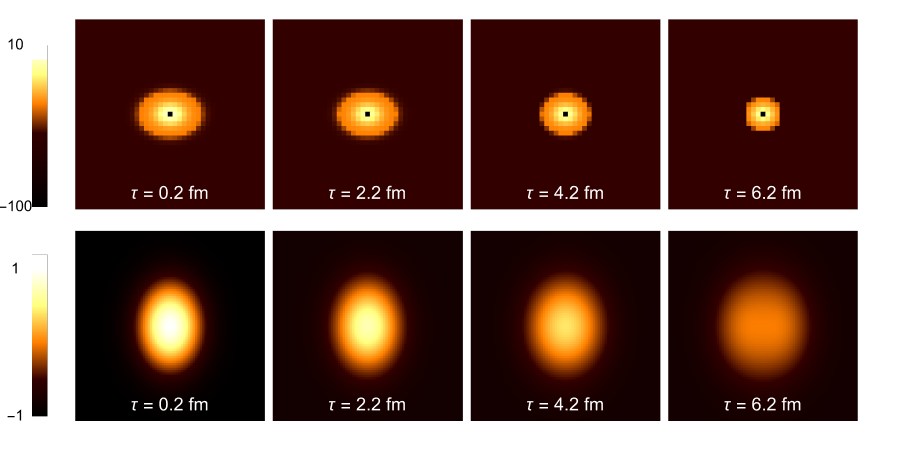}
 \caption{
 The evolution of temperature profile in the transverse $\bm k$-space 
 ($\delta\hat T(\bm k)$ with units $\text{fm}^2$ in the top row) and $\bm x$-space (the dimensionless $\delta\hat T(\bm x)$ in the bottom row)
 at different instants of proper time.
The initial conditions assume the nuclear radius of $6.62$ fm (corresponding to lead). The impact parameter was taken to be $b=6$ fm. The dimensions of the region covered by the images are $2~\text{fm}^{-1}$ by $2~\text{fm}^{-1}$ in $\bm k$-space and  $20$ fm by $20$ fm in $\bm x$-space. }
 \label{fig:gaussian}
\end{figure}

\section{QGP Physics}
\label{sec:observables}

In this section we would like to present some results which indicate that the
set of approximations which we have adopted captures important physical effects
which are beyond the scope of the Bjorken model. In particular, we describe some
quantitative analysis of the numerical calculation described in \rfs{sec:real}.
The numerical solutions \rfn{eq:MIS_finitek} discussed in that section show that
at asymptotically late times the modes representing the perturbations
characterised by nonvanishing wavenumbers are damped away and
eventually the system follows the Bjorken background solution. The QGP never
reaches that stage, however due to hadronization: at some time the system
described by the fluid is converted into a stream of particles. To calculate
their distribution one needs to capture the fine details of the flow during the
freeze-out epoch, when the hydrodynamic variables are converted into a set of
outgoing hadrons over the freeze-out hypersurface $\Sigma$.  This surface will
be taken to be the locus where the local temperature drops below the transition
temperature $T_c \simeq 150$MeV, estimated based on lattice calculations -- we
will be not be distinguishing between chemical and kinetic freeze-out. Within
the linearization framework discussed here, it is reasonable to model freeze-out
at a single instant of proper time, determined by the homogeneous background via
$T(\tau_f)=T_c$. In other words, the freeze-out hypersurface is taken to be both
isothermal and isochronous, the difference between the two being beyond the
linear approximation we consider here. To model the freeze-out surface under
these assumptions we need to introduce its projection onto the transverse plane,
denoted by $\Sigma_\perp$. Its boundary is determined by the condition that
$T(\tau_f, \bm x) = \sigma T_c$ (i.e., $\delta\hat T(\tau_f, \bm x)=\sigma-1$),
where $\sigma$ is a model parameter which is principle could be fitted, but in
this work we choose $\sigma=1/3$. We have found that this leads to reasonable
values of observables such as $v_2$ and $dN/dy$. In the following, the area of
the transverse plane projection of the freeze-out surface $\Sigma_\perp$ is
denoted by $\perparea$.

The approximate determination of the freeze-out surface described above, as well
the fact that the linearization is expected to work less well as the transverse
flow builds up, suggests that one may not be able to obtain quantitative 
agreement with experimental data. We will however see that at least some of the expected
features, such as flow coefficients, can be calculated and are found to be of
the expected order of magnitude.

\subsection{Multiplicity distributions}

In heavy-ion collision experiments,
momentum distributions of hadronized particles are given by the well-known
Cooper–Frye formula~\cite{Cooper:1974}. We first consider the Bjorken
background, with fluid velocity
$u^\mu=(\cosh\eta,\bm 0,\sinh\eta)$, where $\eta$ is the pseudo-rapidity of the
fluid.
For a given particle species with mass $m$ and momentum
$p^\mu=(m_\perp\cosh y,\bm p_\perp, m_\perp\sinh y)$, where
$m_\perp=\sqrt{m^2+\bm p_\perp^2}$, and $y$ is the kinematic rapidity of the
particle, the momentum distribution of particles
created on the freeze-out surface $\Sigma$
is given by
\be
\label{eq:multiplicity_bg_p_phi}
\frac{dN_{\rm B}(p_\perp)}{p_\perp dp_\perp d\phi dy}
=\frac{1}{(2\pi)^3}\int_\Sigma d^3\sigma_\mu p^\mu f(x,p)
=\frac{m_\perp \tau_f\perparea}{(2\pi)^3}F_0.
\ee
Here
\be
    F_0(\tau_f,\hat m,\hat p_\perp)=2K_1(\hat m_\perp)+\frac{1}{12}\left[\hat p_\perp^2 K_1(\hat m_\perp)-2\hat m_\perp K_2(\hat m_\perp)\right]\pa(\tau_f)
    \ee
while $\hat p_\perp\equiv p_\perp/T\equiv|\bm p_\perp|/T$, $\hat m_\perp\equiv
m_\perp/T$, $d^3\sigma_\mu=(\cosh\eta,\bm 0, -\sinh\eta)\tau_f d\eta d^2x$ is
the area element
of the freeze-out surface $\Sigma$,  $f(x,p)=e^{u\cdot\hat
p}(1+\varepsilon_{\mu\nu}\hat p^\mu\hat p^\nu)$ is the nonequilibrium distribution for classical particles with
$\varepsilon_{\mu\nu}=\pi_{\mu\nu}/2(\edens+p)$ being its viscous correction,
and $K_n(x)$ are Bessel functions of the second kind. \rfe{eq:multiplicity_bg_p_phi} reduces
to its NS limit~\cite{Teaney:2003kp} when higher order terms in $\pa$ are neglected.
Note that $\perparea$ cannot be calculated using the
background alone due to the homogeneity in the transverse plane.
Thus, $\perparea$ in \rfe{eq:multiplicity_bg_p_phi} is essentially a regulator and can only be
calculated once the perturbations are included.

At large $\hat p_\perp$, $F_0\sim \frac{\sqrt{2\pi}}{24}\pa\,\hat
p_\perp^{3/2}e^{-\hat p_\perp}$ (cf. the ideal limit where $F_0\sim
\sqrt{2\pi}\hat p_\perp^{-1/2}e^{-\hat p_\perp}$) and thus
\rfe{eq:multiplicity_bg_p_phi} behaves like $\hat p_\perp^{5/2}e^{-\hat p_\perp}$. The
ratio of the dissipative and ideal parts in \rfe{eq:multiplicity_bg_p_phi} is
asymptotically $\pa\,\hat p_\perp^2/24$, which shows that the dissipative part is no longer subleading for sufficiently large $p_\perp$.

\begin{figure}[t]
\centering
\includegraphics[width=.5\textwidth]{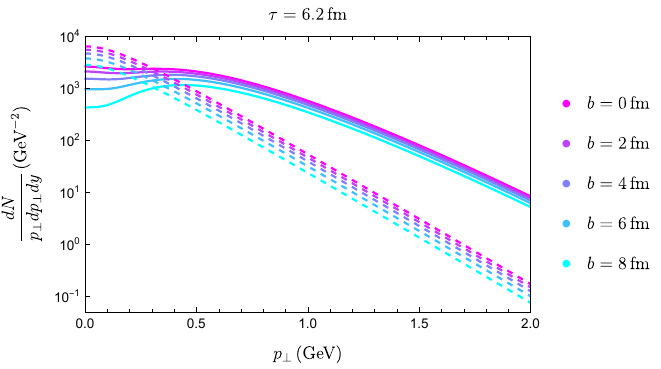}
\hspace{0.02\textwidth}
\includegraphics[width=.4\textwidth]{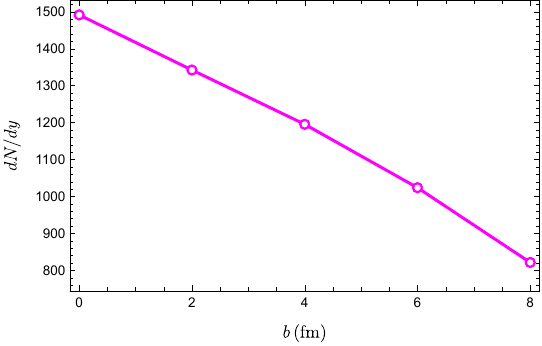}
\caption{This figure shows the dependence of multiplicity distribution on transverse momentum and impact parameter at freeze-out time $\tau=6.2$ fm. The left panel shows the transverse momentum dependence of $\pi^{\pm}$ distribution without and with perturbations (represented by dashed and solid lines respectively). Although the relative contribution from perturbations is more significant compared to its background, its overall magnitude is negligibly smaller compared to the one in the lower $p_\perp$ regime. The right panel shows the impact parameter (centrality) dependence of multiplicity for charged particles including pions, kaons and protons.}
\label{fig:dN/pdpdy}
\end{figure}

Taking into account corrections due to the perturbations up to
quadratic order,
the total multiplicity takes the form
\begin{multline}
\label{eq:multiplicity_pert_p_phi}
\frac{dN(p_\perp,\phi)}{p_\perp dp_\perp d\phi dy}=\frac{m_\perp\tau_f\perparea}{(2\pi)^3}\Big[F_0+F_1\av{\delta\hat T} +F_2\,\hat p_\perp^i\av{\delta u_i}+F_3\,\hat p_\perp^i\hat p_\perp^j\av{\delta\hat\pi_{ij}} +F_{11}\av{\delta\hat T \delta\hat T} \\
+F_{12}\,\hat p_\perp^i\av{\delta u_i\delta\hat T} +F_{13}\,\hat p_\perp^i\hat p_\perp^j\av{\delta\hat\pi_{ij}\delta\hat T}+F_{22}\,\hat p_\perp^i\hat p_\perp^j\av{\delta u_i\delta u_j}+F_{23}\,\hat p_\perp^i\hat p_\perp^j\hat p_\perp^k\av{\delta\hat\pi_{ij}\delta u_k}\Big],
\end{multline}
where $\hat p_{\perp1}=\hat p_\perp\cos\phi$, $\hat p_{\perp2}=\hat
p_\perp\sin\phi$ and $\av{\dots}=\int dx_1dx_2(\dots)/\int dx_1dx_2$ denotes the spacial average over the projected freeze-out surface $\Sigma_\perp$.
The functions $F_1$---$F_{23}$ are given by
\begin{align}\label{eq:F}
    F_1=&~\hat m_\perp\left[K_0(\hat m_\perp)+K_2(\hat m_\perp)\right]-\frac{1}{6}\left[(\hat m_\perp^3+3\hat p_\perp^2)K_1(\hat m_\perp)+3\hat m_\perp(\hat m_\perp-2)K_2(\hat m_\perp)\right]\pa, \nonumber\\
    F_2=&~2F_{22}=F_0=2K_1(\hat m_\perp)+\frac{1}{12}\left[\hat p_\perp^2 K_1(\hat m_\perp)-2\hat m_\perp K_2(\hat m_\perp)\right]\pa, \nonumber\\
    F_3=&~F_{23}=\frac{3}{4}K_1(\hat m_\perp), \nonumber\\
    F_{11}=&~-\hat m_\perp K_0(\hat m_\perp)+\hat m_\perp^2 K_1(\hat m_\perp)+\frac{1}{24}\left[-14\hat m_\perp\hat p_\perp^2K_0(\hat
    m_\perp)\right.\nonumber\\
    &~+\left.2(14\hat m_\perp^2+42\hat p_\perp^2+\hat m_\perp^2\hat p_\perp^2) K_1(\hat m_\perp)-(\hat m_\perp^3+12\hat p_\perp^2) K_2(\hat m_\perp)-\hat m_\perp^3 K_4(\hat m_\perp)\right]\pa, \nonumber\\
    F_{12}=&~2\hat m_\perp K_0(\hat m_\perp)\nonumber\\
    &~+\frac{1}{24}\left[\hat m_\perp(16+\hat p_\perp^2) K_0(\hat m_\perp)+2(16-2\hat m_\perp^2-7\hat p_\perp^2)K_1(\hat m_\perp)+\hat m_\perp\hat p_\perp^2K_2(\hat m_\perp)\right]\pa, \nonumber\\
    F_{13}=&~\frac{3}{4}\left[\hat m_\perp K_0(\hat m_\perp)-5K_1(\hat m_\perp)\right].
\end{align}
In writing these equations we have suppressed the arguments of the functions
$F_i, F_{ij}$. At a conceptual level, \rfe{eq:multiplicity_pert_p_phi} shows
explicitly that all the dependence on transverse dynamics is enters through the
averages of the exponentially suppressed corrections to the homogeneous Bjorken
background.

Integrating \rfes{eq:multiplicity_bg_p_phi} and \rfn{eq:multiplicity_pert_p_phi} over $\phi$ one finds
\be\label{eq:multiplicity_bg_p}
\frac{dN_{\rm B}(p_\perp)}{p_\perp dp_\perp dy}=\frac{m_\perp \tau_f\perparea}{(2\pi)^2}F_0,
\ee
and
\begin{align}\label{eq:multiplicity_pert_p}
\frac{dN(p_\perp)}{p_\perp dp_\perp dy}&=\frac{m_\perp\tau_f\perparea}{(2\pi)^2}\Big[F_0+F_1\av{\delta\hat T}+F_{11}\av{\delta\hat T \delta\hat T} \nonumber\\
&\quad + \frac{1}{2}\hat p_\perp^2\left(F_3\av{\delta\hat\pi_{ii}}+F_{13}\av{\delta\hat\pi_{ii}\delta\hat T}+F_{22}\av{\delta u_i\delta u_i}\right)\Big],
\end{align}
which is proportional to the zeroth order flow coefficient introduced in the
next subsection. At small $\hat p_\perp$, \rfe{eq:multiplicity_pert_p} only
depends on the temperature perturbations. At large $\hat p_\perp$,
\rfe{eq:multiplicity_pert_p} is asymptotically dominated by terms
$\av{\delta\hat T\delta\hat T}$ and $\av{\delta u_i\delta u_i}$, with
coefficients $F_{11}\sim F_{22}\hat p_\perp^2\sim
\frac{\sqrt{2\pi}}{24}\pa\,{\hat p_\perp}^{7/2} e^{-\hat p_\perp}$, so that
\rfe{eq:multiplicity_pert_p} behaves as $\hat p_\perp^{9/2}e^{-\hat
  p_\perp}$. Thus, at large $\hat p_\perp$ the contribution from the
perturbations dominates the background contribution.  Despite this, the
magnitude of the multiplicity distribution in the large $p_\perp$ regime is much
smaller than at small $p_\perp$.  The $p_\perp$ distributions of the total
particle number $N$ given by \rfes{eq:multiplicity_bg_p} and
\rfn{eq:multiplicity_pert_p}, as well as the $p_\perp$-integrated result of
\rfe{eq:multiplicity_pert_p} are shown in \rff{fig:dN/pdpdy}.

The dependence of the multiplicity on both $p_\perp$ and $\phi$ is shown in \rff{fig:dN_p_phi}. In particular, we compare the dependence of the
multiplicity distribution on $\phi$ for two different impact parameters $b$. The
distribution resulting from a central collision ($b=0$) does not depend on
$\phi$, as one would expect, due to the isotropy of the system. On the other
hand, the right panel clearly shows the dependence on the azimuthal angle for a
noncentral collision.  This demonstrates the importance of the contribution from
perturbations for both the magnitude of particle yields and of the angular
distribution in the transverse plane.

\begin{figure}[t]
\centering
\includegraphics[width=1\textwidth]{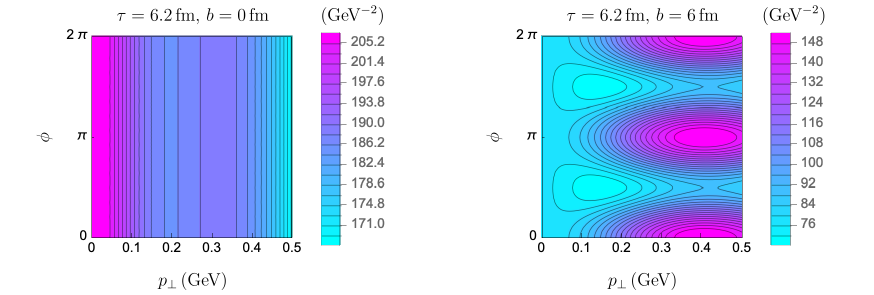}
\caption{This figure shows the 2D contour plot of the multiplicity distribution $dN/p_\perp dp_\perp d\phi dy$ in $(p_\perp,\phi)$ space at $\tau=6.2$ fm with different impact parameters, i.e., $b=0$ fm (left) and $b=6$ fm (right) respectively.}
\label{fig:dN_p_phi}
\end{figure}

\subsection{Collective flow}

In order to study the collective behavior, it is convenient (and standard
practice) to expand the multiplicity distribution geometrically in the
transverse angle $\phi$. This expansion defines a series of flow harmonic
coefficients $v_n$. The coefficients $v_n (p_\perp)$ are defined via the
differential form of the multiplicity distribution in $p_\perp$, i.e.,
\begin{equation}\label{eq:multiplicity_expansion_diff}
    \frac{dN(p_\perp,\phi)}{p_\perp dp_\perp d\phi dy}=v_0(p_\perp)\left(1+\sum_{n=1}^\infty 2v_n(p_\perp)\cos(n\phi)\right),
\end{equation}
while the integrated ones are defined via the multiplicity distribution after its integration over $p_\perp$, i.e.,
\begin{equation}\label{eq:multiplicity_expansion_int}
    \frac{dN(\phi)}{d\phi dy}=v_0\left(1+\sum_{n=1}^\infty 2v_n\cos(n\phi)\right).
\end{equation}
From \rfe{eq:multiplicity_expansion_diff} and \rfe{eq:multiplicity_expansion_int}, one finds
\begin{align}\label{eq:v0}
    v_0(p_\perp)&=\int_0^{2\pi}\frac{dN(p_\perp,\phi)}{p_\perp dp_\perp d\phi dy}\frac{d\phi}{2\pi}=\frac{1}{2\pi}\frac{dN(p_\perp)}{p_\perp dp_\perp dy}, \nonumber\\ v_0&=\int_0^{2\pi}\frac{dN(\phi)}{d\phi dy}\frac{d\phi}{2\pi}=\frac{1}{2\pi}\frac{dN}{dy}=\int_0^\infty dp_\perp p_\perp v_0(p_\perp).
\end{align}
That said, $v_0(p_\perp)$ is the multiplicity averaged over the transverse angle
$\phi$, while $v_0$ is the total multiplicity $dN/dy$, up to a factor of
$1/2\pi$. Anisotropy of the flow is reflected by the nonvanishing of the
coefficients $v_n$ for $n>0$.  From \rfe{eq:multiplicity_expansion_diff}
and \rfe{eq:multiplicity_expansion_int} one can also identify
\begin{align}\label{eq:vn}
    v_n(p_\perp)&=\frac{1}{v_0(p_\perp)}\int_0^{2\pi}\frac{dN(p_\perp,\phi)}{p_\perp dp_\perp d\phi dy}\frac{d\phi}{2\pi}\cos(n\phi), \nonumber\\
    v_n&=\frac{1}{v_0}\int_0^{2\pi}\frac{dN(\phi)}{d\phi dy}\frac{d\phi}{2\pi}\cos(n\phi)=\frac{\int_0^\infty dp_\perp p_\perp v_0(p_\perp)v_n(p_\perp)}{\int_0^\infty dp_\perp p_\perp v_0(p_\perp)}.
\end{align}
The last equality in \rfe{eq:vn} shows that the integrated coefficients $v_n$ represent the averaged values of $v_n(p_\perp)$ over the momentum spectrum.

In this paper we focus on the first few harmonic coefficients of $v_n$ when $n=0,1,2$, which describe  the direct, radial, and elliptic flow respectively.
Substituting \rfe{eq:multiplicity_pert_p_phi} into \rfe{eq:vn}, one immediately finds
\begin{subequations}\label{eq:v0123pT}
\begin{align}
    v_0(\hat p_\perp)&=\frac{m_\perp\tau_f\perparea}{(2\pi)^3}\bigg[F_0+F_1\av{\delta\hat T}+F_{11}\av{\delta\hat T\delta\hat T}\nonumber\\
    &\quad+\frac{1}{2}\hat p_\perp^2\left(F_3\av{\delta\hat\pi_{ii}}+F_{13}\av{\delta\hat T\delta\hat\pi_{ii}}+F_{22}\av{\delta u_i\delta u_i}\right)\bigg],\\
    v_1(\hat p_\perp)&=\frac{m_\perp\tau_f\perparea\hat p_\perp}{2(2\pi)^3v_0(\hat p_\perp)}\bigg[F_2\av{\delta u_1}+F_{12}\av{\delta\hat T\delta u_1}\nonumber\\
    &\quad+\frac{1}{4}F_{23}\hat p_\perp^2(\av{\delta u_1(3\delta\hat\pi_{11}+\delta\hat\pi_{22})}+2\av{\delta u_2\delta\hat\pi_{12}})\bigg],\\
    v_2(\hat p_\perp)&=\frac{m_\perp\tau_f\perparea\hat p_\perp^2}{4(2\pi)^3v_0(\hat p_\perp)}\left[F_3\av{\delta\hat\pi_{11}-\delta\hat\pi_{22}}+F_{13}\av{\delta\hat T(\delta\hat\pi_{11}-\delta\hat\pi_{22})}+F_{22}\av{\delta u_1^2-\delta u_2^2}\right],
\end{align}
\end{subequations}
where we suppressed the argument $p_\perp$ of the coefficient functions $F_i,
F_{ij}$ given by \rfe{eq:F}. The asymptotic behavior of $v_0(p_\perp)$, $v_1(p_\perp)$, and $v_2(p_\perp)$ at small momentum is found to be
\begin{subequations}
\begin{gather}
    v_0(p_\perp)\sim 1, \quad v_1(p_\perp)\sim p_\perp, \quad v_2(p_\perp)\sim p_\perp^2 \quad\text{as}\quad \hat p_\perp\to0,
\end{gather}
while at large momentum one has
\begin{gather}
    v_0(p_\perp)\sim \hat p_\perp^{9/2}e^{-\hat p_\perp}, \quad v_1(p_\perp)\sim \hat p_\perp^{-1}, \quad v_2(p_\perp)\sim\frac{\av{\delta u_1^2-\delta u_2^2}}{2(4\av{\hat\delta T\delta\hat T}+\av{\delta u_i\delta u_i})}\sim 1 \quad\text{as}\quad \hat p_\perp\to\infty.
\end{gather}
\end{subequations}
The asymptotic behavior of $v_0(p_\perp)$ has already been analyzed below \rfe{eq:multiplicity_pert_p} where we studied  the $\phi$-integrated
multiplicity which is related to $v_0(p_\perp)$ via \rfe{eq:v0}. We
observe that higher-order flow harmonics grow faster than $p_\perp$ for small
$p_\perp$ and decrease slower at large $p_\perp$. In particular, we find that
$v_2(p_\perp)$ increases quadratically at small $p_\perp$, and saturates at
large $p_\perp$. The dependence of the elliptic flow on the transverse momentum
for various collision centralities (impact parameters) is illustrated in
\rff{fig:v2}. 
The direct flow $v_1$, on the other hand, vanishes due to the
specific symmetric initial condition which we used, which means that in this
case, on average, there is no preferred
direction for particle emission.

\begin{figure}[t]
\includegraphics[width=0.5\textwidth]{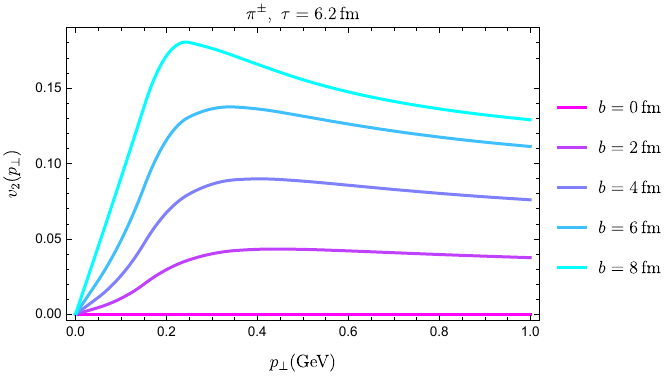}
\hspace{0.02\textwidth}
\includegraphics[width=.4\textwidth]{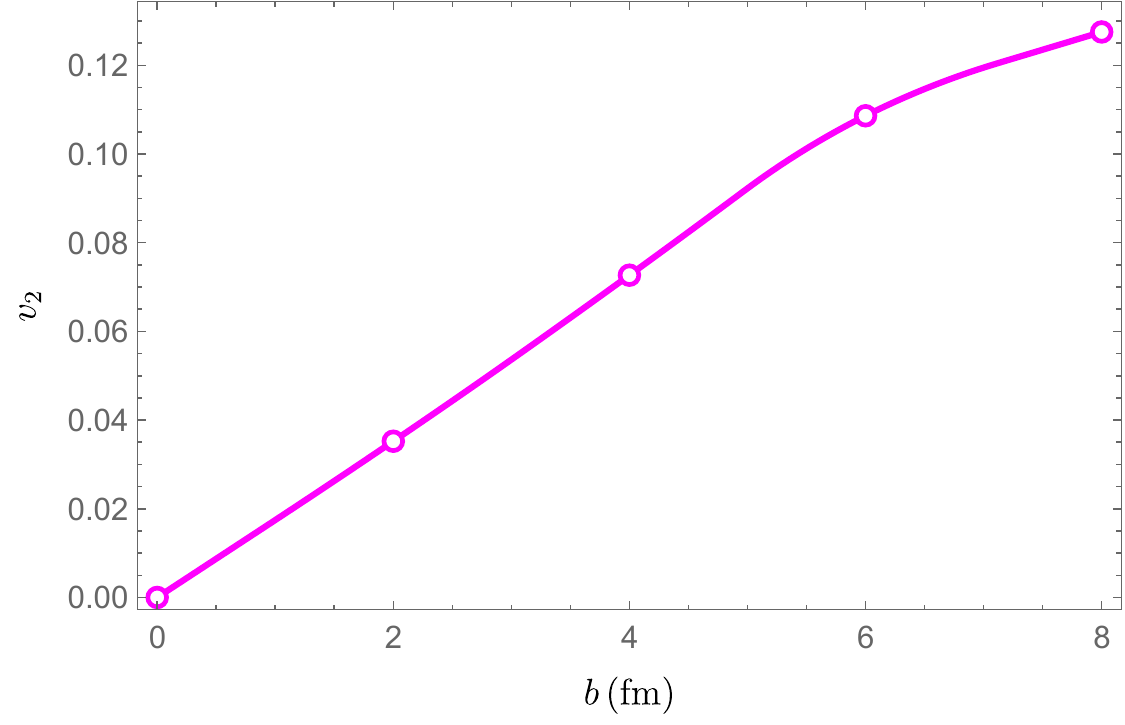}
\caption{This figure shows the dependence of elliptic flow on transverse momentum and impact parameter at freeze-out time $\tau=6.2$ fm. The left panel shows the dependence of differential $v_2$ on transverse momentum for pions, while the right panel shows the dependence of the integrated $v_2$ on impact parameter.}
    \label{fig:v2}
\end{figure}

The integrated coefficients $v_n$ can be obtained using \rfes{eq:vn} and \rfn{eq:v0123pT}. Focusing on $v_2$ we find the following analytic expression
\begin{subequations}\label{eq:v2_int}
\begin{multline}
    v_2=\frac{T^3\tau\perparea}{v_0(2\pi)^3}\Bigg\{-\frac{3}{16}(\hat m^2\mathcal I_{2,1}-\mathcal I_{4,1})\av{\delta\hat\pi_{ii}} -\frac{3}{16}\left[\hat m^2(5\mathcal I_{2,1}-\mathcal I_{3,0})-5\mathcal I_{4,1}+\mathcal I_{5,0}\right]\av{\delta\hat T\delta\hat\pi_{ii}}\\
    +\left[\frac{1}{4}(-\hat m^2\mathcal I_{2,1}+\mathcal I_{4,1})+\frac{\pa}{96}(\hat m^4\mathcal I_{2,1}+2\hat m^2({\mathcal I_{3,2}-\mathcal I_{4,1}})-2\mathcal I_{5,2}+\mathcal I_{6,1})\right]\av{\delta u_i\delta u_i}\Bigg\}
\end{multline}
\text{where}
\begin{multline}
    v_0=\frac{T^3\tau\perparea}{(2\pi)^3}\Bigg\{2I_{2,1}-\frac{\pa}{12}(\hat m^2\mathcal I_{2,1}+2\mathcal I_{3,2}-\mathcal I_{4,1})\\
    +\left[\mathcal I_{3,0}+\mathcal I_{3,2}+\frac{\pa}{6}(3\hat m^2\mathcal I_{2,1}+6\mathcal I_{3,2}-3\mathcal I_{4,1}-3\mathcal I_{4,2}-\mathcal I_{5,1})\right]\av{\delta\hat T}\\
     -\frac{3}{8}(\hat m^2\mathcal I_{2,1}-\mathcal I_{4,1})\av{\delta\hat\pi_{ii}} -\frac{3}{8}\left[\hat m^2(5\mathcal I_{2,1}-\mathcal I_{3,0})-5\mathcal I_{4,1}+\mathcal I_{5,0}\right]\av{\delta\hat T\delta\hat\pi_{ii}}\\
    -\bigg[\mathcal I_{3,0}-\mathcal I_{4,1}\\
    +\frac{\pa}{24}\left(2\hat m^2(42\mathcal I_{2,1}-6\mathcal I_{2,2}-7\mathcal I_{3,0}+\mathcal I_{4,1})-112\mathcal I_{4,1}+12\mathcal I_{4,2}+14\mathcal I_{5,0}+\mathcal I_{5,2}+\mathcal I_{5,4}-2\mathcal I_{6,1}\right)\bigg]\av{\delta\hat T^2}\\
    +\left[\frac{1}{2}(-\hat m^2\mathcal I_{2,1}+\mathcal I_{4,1})+\frac{\pa}{48}(\hat m^4\mathcal I_{2,1}+2\hat m^2({\mathcal I_{3,2}-\mathcal I_{4,1}})-2\mathcal I_{5,2}+\mathcal I_{6,1})\right]\av{\delta u_i\delta u_i}\Bigg\}.
\end{multline}
\end{subequations}
The functions $\mathcal I_{\ell,n}(\hat m)$ are integrals defined in
Appendix~\ref{app:Integrals}, where they are expressed in terms of known special functions.

\section{Summary and Outlook}
\label{sec:outlook}

Despite the radical simplifications made, the original Bjorken model of
Ref.~\cite{Bjorken:1982qr} provided a very useful analytic formula describing
the dynamics of the energy density and has lead to many important insights. The
subsequent incorporation of viscous effects and higher orders in the large
proper time expansion retained some of this simplicity, 
allowing for approximate analytic calculations and significantly
simplifying numerical computations. However, the stringent symmetry requirements
-- boost invariance and homogeneity in the transverse plane -- placed severe
limits on how much of the physics could be described  by such models.
In this article our goal was to generalize the Bjorken model by incorporating
the transverse dynamics in an approximate way so as to capture physically
crucial effects such as elliptic flow, which are inaccessible
as long as all the symmetries of the Bjorken model are maintained.
We have therefore partially relaxed these symmetry requirements, insisting only
on longitudinal boost invariance while linearizing around a boost-invariant,
transversely homogeneous attractor background, characterized only by a choice of
the initial temperature. This approach aims to balance simplicity with the
ability to capture more of the interesting physics of QGP.
In this way we have formulated a description which goes beyond toy models, but
still allows some analytic insights which can be extracted by applying asymptotic techniques.

The initial state is encoded in the initial conditions for the Fourier modes of
the transverse perturbations and the initial temperature of the attractor
background. The evolution equations map this information into a set of six
exponentially suppressed scale-dependent amplitudes for each Fourier mode and
the scale $\Lambda$. From the perspective of modern asymptotic analysis our work
shows that the dynamics of QGP created in heavy-ion collision experiments
provides a physical situation where the exponentially suppressed corrections to
asymptotic power series solutions carry almost all the information which is
actually detected in experiments. We have shown how these exponential
corrections are translated into the physics post-freeze-out.

Aside from providing a clear, semianalytic picture of QGP evolution, our
approach can be implemented numerically in a very straightforward and efficient
fashion, since it relies on the discrete Fourier transforms and solving systems
of coupled linear ODEs. The details of the initial state can be incorporated
scale by scale in a controlled way: to account for finer details of the initial
state, one can increase the number of modes used in the calculation.  In this
pilot study we have considered the build-up of elliptic flow and have found
results qualitatively consistent with earlier work which relied on solving the
fully-nonlinear problem (e.g. Ref.~\cite{Luzum:2008cw}). It is not clear how
much quantitative agreement with actual experimental data can be achieved along
these lines, one reason being that the applicability of linearization around the
attractor is expected to deteriorate in the course of evolution due to the
growth of transverse flow and may be somewhat rough by the time of freeze-out.
It would be very interesting to better understand the limitations of our
approach, identifying which physical effects are captured by this model, and
which require a fully nonlinear treatment -- an example of the latter may be
turbulent phenomena~\cite{Floerchinger:2011pxy}.

Our approach, apart from its conceptual aspects, could also be used as a laboratory
for studying models of the initial state, as well as novel characterizations of
fluid behavior~\cite{Ollitrault:2023wjk,Ambrus:2022qya}. Our model can also be
extended in a number of ways; two most prominent are the description of jets
\cite{Chaudhuri_2006,Chesler:2007an,Casalderrey-Solana:2020rsj,Yang_2023}, and
the incorporation of noise
\cite{Akamatsu:2017,An:2019rhf,An:2019fdc,An_2021,An_2023,Chen:2022ryi}. We hope
to return to these matters in the future.

\section{Acknowledgments}

It is a pleasure to thank V. Ambrus, L. Du, W. Florkowski, S. Mrówczyński and M.
Stephanov for helpful discussions. The authors would also like to thank the Isaac Newton Institute for Mathematical Sciences,
Cambridge, for support and hospitality during the program “Applicable resurgent
asymptotics: towards a universal theory” supported by EPSRC Grant No.
EP/R014604/1. This work was supported by the National Science Centre, Poland, under Grant No. 2021/41/B/ST2/02909. For the purpose of Open Access, the authors have applied a
CC-BY public copyright licence to any Author Accepted Manuscript (AAM) version arising from this submission.

\section*{Appendix}

\appendix

\section{Linearized ODEs}
\label{app:ODE}

The six first-order linear differential equations in \rfes{eq:linsys2} can be written as three second-order linear differential equations by eliminating the variables $\pi_{ij}$, the resulting equations take the form
\begin{subequations}\label{eq:mispert2}
\begin{align}
  \partial^2_\tau\delta\hat T+P^{(1)}_{T T}\partial_\tau\delta\hat T(\tau)+P^{(1)}_{T \theta}\partial_\tau\delta\hat\theta+P^{(0)}_{T T}\delta\hat T+P^{(0)}_{T \theta}\delta\hat\theta=0,\\
  \partial^2_\tau\delta\hat\theta+P^{(1)}_{\theta T}\partial_\tau\delta\hat T+P^{(1)}_{\theta \theta}\partial_\tau\delta\hat\theta+P^{(0)}_{\theta T}\delta\hat T+P^{(0)}_{\theta \theta}\delta\hat\theta=0,\\
\partial^2_\tau\delta\hat\omega+P^{(1)}_{\omega\omega}\partial_\tau\delta\hat\omega+P^{(0)}_{\omega\omega}\delta\hat\omega=0,
\end{align}
\end{subequations}
with coefficients $P_{AB}^{(n)}$, where $A,B=T,\theta,\omega$ and $(n)$ denotes the differentiation order of the corresponding term:
\begin{align}
P^{(1)}_{TT}=&~\frac{9(1+w)+4\pa}{9\tau}, \nonumber\\
P^{(1)}_{T\theta}=&~\frac{k(12+\pa)}{36},\nonumber\\
P^{(0)}_{TT}=&~\frac{-288\alpha^2+\left(45w+8\pa+36\tau\partial_\tau\right)\pa}{162\tau^2},\nonumber\\
P^{(0)}_{T\theta}=&~\frac{k}{324\tau}\left[36(3+2\alpha^2+3w)+(57+9w+2\pa+9\tau\partial_\tau)\pa\right],
\nonumber\\
P^{(1)}_{\theta T}=&~\frac{12k}{12+\pa},\nonumber\\
P^{(1)}_{\theta\theta}=&~P^{(1)}_{\omega\omega}=\frac{-36(1-2\alpha^2-3w)+(45+2\pa+9\tau\partial_\tau)\pa}{9(12+\pa)\tau},\nonumber\\
P^{(0)}_{\theta T}=&~-\frac{k[12(8\alpha^2+3w)+\left(8-3w\right)\pa]}{3(12+\pa)\tau},\nonumber\\
P^{(0)}_{\theta\theta} =&~ \frac{1}{27(12+\pa)\tau^2}\left\{108\left[(1-2\alpha^2)(1-w)+4\alpha^2k^2\tau^2\right] +\left[-\frac{3}{2}(58-32\alpha^2-6w+w\pa)\right.\right.\nonumber\\
&~\left.\left. +36k^2\tau^2+(7-3w)(9w+2\pa+9\tau\partial_\tau)\right]\pa\right\},\nonumber\\
P^{(1)}_{\omega\omega}=&~(7-3w)\left(9w+2\pa+9\tau\partial_\tau\right)\pa,\nonumber\\
P^{(0)}_{\omega\omega}=&~\frac{1}{27(12+\pa)
\tau^2}\left\{108\left[(1-2\alpha^2)(1-w)+3\alpha^2k^2\tau^2\right]\right.\nonumber\\
&~\left.+\left[-\frac{3}{2}(58-32\alpha^2-6w+w\pa)+(7-3w)\left(9w+2\pa+9\tau\partial_\tau\right)\right]\pa\right\},
\end{align}
where $w\equiv\tau/\tau_\pi=\tau T/C_\tau$. One immediately notices that the equation for vorticity $\delta\hat\omega$ in \rfes{eq:mispert2} decouples from the other two equations. Given a solution to these equations, one can recover the independent shear-stress tensor components algebraically.

Substituting \rfe{eq:nsbjA} into \rfes{eq:mispert2}, and taking the NS limit $\tau_\pi\to0$ (i.e.,  $C_\tau\to0$), one obtains
\begin{subequations}\label{eq:linsys2NS}
\begin{align}
    \left(\partial_\tau+\frac{4C_\eta}{9\tau^2 T}\right)\delta\hat T+\frac{k}{3}\left(1+\frac{4C_\eta}{3\tau T}\right)\delta\hat\theta&=0,\\
    -\frac{3(2C_\eta+\tau T)}{2C_\eta+3\tau T}k\delta\hat T+\left(\partial_\tau+\frac{2C_\eta(7+6k^2\tau^2)-3\tau T}{3\tau(2C_\eta+3\tau T)}\right)\delta\hat\theta&=0,\\
    \left(\partial_\tau+\frac{C_\eta(14+9k^2\tau^2)-3\tau T}{3\tau(2C_\eta+3\tau T)}\right)\delta\hat\omega&=0.
\end{align}
\end{subequations}
Note that in these equations $T(\tau)$ is the background solution given explicitly in \rfe{eq:nsbjT}.
The above equations can also be obtained, by noting that in the NS limit the shear stress tensor perturbations $\delta\hat\pi_{ij}$ are no longer independent degrees of freedom, and hence can be expressed in terms of the hydrodynamic variables via
\begin{align}\label{eq:dpi_NS}
    \delta\hat\pi_{ij}(\tau,\bm k)=\frac{1}{3}\left[\delta_{ij}\delta\hat T+\left(\frac{1}{3}(\delta_{ij}-3\hat k_i\hat k_j)\delta\hat\theta(\tau,\bm k)+\frac{1}{2}(\hat k_i\epsilon_{j\ell}+\hat k_j\epsilon_{i\ell})\hat k^\ell\delta\hat\omega(\tau,\bm k)\right)k\tau\right]\pa_{\rm NS}(\tau),
\end{align}
which is resulted from \rfe{eq:linsys2-4} by setting $C_\tau=0$ and $\pa=\pa_{\rm NS}$. Substituting the above expression into \rfes{eq:linsys2} and using again $\pa=\pa_{\rm NS}$, one reproduces \rfes{eq:linsys2NS}.

The ideal fluid equations are  obtained simply by setting $C_\eta=0$ in \rfes{eq:linsys2NS}.

We also note that \rfe{eq:linsys2NS} can be approximated further 
in the late-time limit, when the timescale of the system is much larger than the typical microscopic timescale characterized by the shear viscosity: $\tau\gg\tau_{\rm mic}\sim\eta/(\edens+p)$, or  equivalently $C_\eta/\tau T\ll1$. In this case, \rfes{eq:linsys2NS} reduce to the equations studied in \cite{Floerchinger:2011pxy}.

\section{The late-time asymptotic solutions in Navier-Stokes and ideal limits}
\label{app:NS-ID}
The asymptotic solutions in the NS limit ($C_{\tau}=0$) can be obtained using \rfes{eq:linsys2NS}. In the long-time limit $\tau\to\infty$, we find
\begin{subequations}
\begin{align}
    \delta\hat T_{\rm NS}(\tau)&\sim C_1(\Lambda\tau)^{-\beta_4^{\rm NS}}e^{-\frac{C_\eta k^2}{\Lambda^2}(\Lambda\tau)^{\frac{4}{3}}+\frac{3}{8C_\eta}(\Lambda\tau)^{\frac{2}{3}}}\left(1+\mathcal O((\Lambda\tau)^{-\frac{2}{3}})\right)\nonumber\\
    &\quad +C_2(\Lambda\tau)^{\beta_4^{\rm NS}}e^{-\frac{3}{8C_\eta}(\Lambda\tau)^{\frac{2}{3}}}\left(1+\mathcal O((\Lambda\tau)^{-\frac{2}{3}})\right),\\
    \delta\hat\theta_{\rm NS}(\tau)&\sim C_1\frac{4C_\eta k}{\Lambda}(\Lambda\tau)^{-\beta_4^{\rm NS}+\frac{1}{3}}e^{-\frac{C_\eta k^2}{\Lambda^2}(\Lambda\tau)^{\frac{4}{3}}+\frac{3}{8C_\eta}(\Lambda\tau)^{\frac{2}{3}}}\left(1+\mathcal O((\Lambda\tau)^{-\frac{2}{3}})\right)\nonumber\\
    &\quad+C_2\frac{3\Lambda}{4C_\eta k}(\Lambda\tau)^{\beta_4^{\rm NS}-\frac{1}{3}}e^{-\frac{3}{8C_\eta}(\Lambda\tau)^{\frac{2}{3}}}\left(1+\mathcal O((\Lambda\tau)^{-\frac{2}{3}})\right),\\
    \delta\hat\omega_{\rm NS}(\tau)&\sim C_3(\Lambda\tau)^{1/3}e^{-\frac{3C_\eta k^2}{4\Lambda^2}(\Lambda\tau)^{\frac{4}{3}}}\left(1+\frac{8C_\eta}{3}(\Lambda\tau)^{-2/3}+O((\Lambda\tau)^{-\frac{4}{3}})\right),
\end{align}
\end{subequations}
 where $\beta_4^{\rm NS}=-\frac{2}{3}-\frac{3\Lambda^2}{64C_\eta^3k^2}$.
In the ideal limit, $C_{\tau}=C_\eta=0$, from \rfes{eq:linsys2NS}
one immediately obtains~\cite{Floerchinger:2011pxy}
\begin{subequations}
\begin{align}
    \delta\hat T_{\rm I}(\tau)&=(\Lambda\tau)^{1/6}\left(C_1e^{-\frac{\mathrm i}{\sqrt{3}}k\tau}+C_2e^{\frac{\mathrm i}{\sqrt{3}}k\tau}\right), \\
    \delta\hat\theta_{\rm I}(\tau)&=\mathrm i\sqrt{3}(\Lambda\tau)^{1/6}\left(C_1e^{-\frac{\mathrm i}{\sqrt{3}}k\tau}-C_2e^{\frac{\mathrm i}{\sqrt{3}}k\tau}\right),\\
    \delta\hat\omega_{\rm I}(\tau)&=C_3(\Lambda\tau)^{1/3},
\end{align}
\end{subequations}
where the coefficients $\{C_n(k)\}$ are integration constants.

It is interesting to consider $k=0$ solutions in \rfe{eq:MIS_zerok} in the NS limit $C_\tau\rightarrow 0$ and compare them to late-time solutions of the NS \rfes{eq:linsys2NS}, which read
\begin{align}\label{eq:zerok_NS}
    \delta  u_i^{\rm NS}(\tau)&\sim C_i(\Lambda\tau)^{\frac{1}{3}}\left(1+\frac{8C_\eta}{3}(\Lambda\tau)^{-\frac{2}{3}}+\frac{32C_\eta^2}{9}(\Lambda\tau)^{-\frac{4}{3}}+\mathcal O((\Lambda\tau)^{-2})\right),\nonumber\\
    \delta{\hat T}^{\rm NS}(\tau)&\sim C_3 \left(1+\frac{2C_\eta}{3}(\Lambda\tau)^{-\frac{2}{3}}+\frac{4C_\eta^2}{9}(\Lambda\tau)^{-\frac{4}{3}}+\mathcal O((\Lambda\tau)^{-2})\right), \nonumber\\
    \delta\hat\pi_{ii}^{\rm NS}(\tau)&\sim C_3\frac{8C_\eta}{3}(\Lambda\tau)^{-\frac{2}{3}}\left(1+\frac{4C_\eta}{3}(\Lambda\tau)^{-\frac{2}{3}}+\mathcal O((\Lambda\tau)^{-\frac{4}{3}}\right), \quad \delta\hat\pi_{12}^{\rm NS}(\tau)=0.
\end{align}
The exponential contributions to the solutions of $\delta\hat T$ and $\delta\hat\pi_{ij}$ in \rfe{eq:MIS_zerok} vanish in this limit, the remaining asymptotic solutions are related via \rfe{eq:dpi_NS}. While the solution of $\delta{\hat{T}}$ and $\delta\hat\pi_{ij}$ coincides with the $C_\tau\rightarrow 0$ limit of the MIS result in \rfe{eq:MIS_zerok}, the series for $\delta u_i^{\rm NS}$ deviates from the corresponding MIS result at subleading orders. This is a manifestation of the fact that the late-time expansion does not necessarily commute with the $\tau_\pi\rightarrow 0$ limit.

The solution in the ideal limit can be obtained by taking $C_\eta=0$ in \rfe{eq:zerok_NS}. We note that the $\tau^{1/3}$ growth of the $k=0$ velocity mode is a common feature shared by MIS, NS and perfect-fluid hydrodynamics.
Indeed, one can see directly from the $k=0$ perfect-fluid equations that the decrease of the energy density caused by the longitudinal expansion along with the conservation of transverse momentum imply that the velocity perturbation much grow is a way which compensates the aforementioned fall-off of the background energy density.
This suggests it is a feature of perturbations of flows satisfying the Bjorken symmetry assumptions and likely affects homogeneous perturbations of such flows regardless of the dynamics.

\section{Integrals for calculating elliptic flow}
\label{app:Integrals}

The functions $\mathcal I_{\ell,n}(\hat m)$ appearing in \rfes{eq:v2_int} are defined by
\begin{equation}
    \mathcal I_{\ell,n}(\hat m)\equiv\int_{\hat m}^\infty dx x^\ell K_n(x).
\end{equation}
This integration results from changing the integration variable from $\hat p_\perp$ to $x=\hat m_\perp=\sqrt{\hat m^2+\hat p^2}$, where $\hat m\equiv m/T$. The analytic expression of integrals what needed for calculating $v_2$ are
\begin{gather}
    \mathcal I_{3,0}=4G_{1,3}^{3,0}\left(
    \begin{array}{c}
    1 \\
    0,2,2 \\
    \end{array}\Big|\frac{\hat m^2}{4}\right), \quad
    \mathcal I_{5,0}=16G_{1,3}^{3,0}\left(
    \begin{array}{c}
    1 \\
    0,3,3 \\
    \end{array}\Big|\frac{\hat m^2}{4}\right), \quad
    \mathcal I_{0,1}=K_0(\hat m), \nonumber\\
    \mathcal I_{2,1}=\hat m^2K_2(\hat m), \quad \mathcal I_{4,1}=8 G_{1,3}^{3,0}\left(
\begin{array}{c}
 1 \\
 0,2,3 \\
\end{array}\Big|\frac{\hat m^2}{4}
\right), \quad \mathcal I_{6,1}=32 G_{1,3}^{3,0}\left(
\begin{array}{c}
 1 \\
 0,3,4 \\
\end{array}\Big|\frac{\hat m^2}{4}
\right), \nonumber\\
\mathcal I_{5,1}=-\frac{1}{5} \hat m^6 K_1(\hat m) \, _1F_2
\left(\begin{array}{c}
1 \\
\frac{7}{2},\frac{7}{2}\\
\end{array}\Big|\frac{\hat m^2}{4}\right)
-\hat m^5 K_0(\hat m) \left(\, _1F_2
\left(\begin{array}{c}
1\\
\frac{5}{2},\frac{7}{2}\\
\end{array}\Big|\frac{\hat m^2}{4}\right)-1\right)+\frac{45\pi}{2},\nonumber\\
\mathcal I_{3,2}=\hat m^3K_3(\hat m), \quad
\mathcal I_{5,2}=16 G_{1,3}^{3,0}\left(
\begin{array}{c}
 1 \\
 0,2,4 \\
\end{array}\Big|\frac{\hat m^2}{4}
\right), \quad
\mathcal I_{5,4}=\hat m^5K_5(\hat m),\nonumber\\
\mathcal I_{2,2}=-\frac{\hat m^5}{200}\left\{5\left[\partial_x{}_2F_3
\left(\begin{array}{c}
\frac{5}{2},x\\
1,3,\frac{7}{2}\\
\end{array}\Big|\frac{\hat m^2}{4}\right)\right]_{x=1}+{}_2F_3\left(\begin{array}{c}
\frac{5}{2},\frac{5}{2}\\
3,\frac{7}{2},\frac{7}{2}\\
\end{array}\Big|\frac{\hat m^2}{4}\right)\right\}\nonumber\\
+\frac{\hat m^5}{120}(3\log \hat m+3\gamma+4-\log8) \, _1F_2\left(\begin{array}{c}
\frac{5}{2}\\
3,\frac{7}{2}\\
\end{array}\Big|\frac{\hat m^2}{4}\right)+\frac{\hat m^3}{3}-\frac{4 \hat m^2 I_1(\hat m)}{3}-4 \hat m+2 \hat m I_0(\hat m)+\frac{3\pi}{2},\nonumber\\
\mathcal I_{4,2}=-\frac{1}{392}\hat m^7 \left\{7\partial_x\left[{}_2F_3\left(\begin{array}{c}
\frac{7}{2},x\\
1,3,\frac{9}{2}\\
\end{array}\Big|\frac{\hat m^2}{4}\right)\right]_{x=1}+{}_2F_3\left(\begin{array}{c}
\frac{7}{2},\frac{7}{2}\\
3,\frac{9}{2},\frac{9}{2}\\
\end{array}\Big|\frac{\hat m^2}{4}\right)\right\}\nonumber\\
+\frac{1}{840} \hat m^7 (15 \log \hat m+15 \gamma +8-15 \log2) \, _1F_2\left(\begin{array}{c}
\frac{7}{2}\\3,\frac{9}{2}\\
\end{array}\Big|\frac{\hat m^2}{4}\right)\nonumber\\
+\frac{\hat m^5}{5}-\frac{8 \hat m^4 I_1(\hat m)}{15}-\frac{4 \hat m^3}{3}+\frac{2 \hat m^3 I_0(\hat m)}{3}+\frac{15\pi}{2},
\end{gather}
where several new special functions, in addition to the gamma function $\Gamma(x)$ and the Bessel function of the second kind $K_n(x)$ discussed above, are introduced:
\begin{equation}
    G^{m,n}_{p,q}
\left(
\begin{array}{c}
 \bm{a_p} \\
 \bm{b_q} \\
\end{array}
\Big| z
\right)=\frac{1}{2\pi i}\int_L\frac{\prod_{j=1}^n\Gamma(1-a_j+s)\prod_{j=1}^m\Gamma(b_j-s)}{\prod_{j=n+1}^p\Gamma(a_j-s)\prod_{j=m+1}^q\Gamma(1-b_j+s)}
\end{equation}
is the Meijer G-function represented in the Mellin–Barnes-type line integral and $\bm a_p=(a_1,\dots,a_p)$, $\bm b_p=(b_1,\dots,b_p)$.
\be
{}_pF_q
\left(
\begin{array}{c}
 \bm{a_p} \\
 \bm{b_q} \\
\end{array}
\Big| z
\right)
= \sum_{n=0}^{\infty} \frac{(a_1)_n\dots(a_p)_n}{(b_1)_n\dots (b_q)_n} \frac{z^n}{n!}, \qquad (a)_n = \frac{\Gamma(a+n)}{\Gamma(a)}=a(a+1)(a+2) \ldots (a+n-1)
\ee
is the generalized hypergeometric function. $I_n(z)$ is the  modified Bessel function of the first kind.

\bibliographystyle{JHEP}
\bibliography{transverse}{}

\providecommand{\href}[2]{#2}\begingroup\raggedright\begin{thebibliography}{10}

\bibitem{Bjorken:1982qr}
J.D.~Bjorken, \emph{{Highly Relativistic Nucleus-Nucleus Collisions: The
  Central Rapidity Region}},
  \href{https://doi.org/10.1103/PhysRevD.27.140}{\emph{Phys. Rev. D} {\bfseries
  27} (1983) 140}.

\bibitem{Heller:2015dha}
M.P.~Heller and M.~Spaliński, \emph{{Hydrodynamics Beyond the Gradient
  Expansion: Resurgence and Resummation}},
  \href{https://doi.org/10.1103/PhysRevLett.115.072501}{\emph{Phys. Rev. Lett.}
  {\bfseries 115} (2015) 072501}
  [\href{https://arxiv.org/abs/1503.07514}{{\ttfamily 1503.07514}}].

\bibitem{Romatschke:2017vte}
P.~Romatschke, \emph{{Relativistic Fluid Dynamics Far From Local Equilibrium}},
  \href{https://doi.org/10.1103/PhysRevLett.120.012301}{\emph{Phys. Rev. Lett.}
  {\bfseries 120} (2018) 012301}
  [\href{https://arxiv.org/abs/1704.08699}{{\ttfamily 1704.08699}}].

\bibitem{Kurkela_2020}
A.~Kurkela, W.~van~der Schee, U.A.~Wiedemann and B.~Wu, \emph{{Early- and
  Late-Time Behavior of Attractors in Heavy-Ion Collisions}},
  \href{https://doi.org/10.1103/PhysRevLett.124.102301}{\emph{Phys. Rev. Lett.}
  {\bfseries 124} (2020) 102301}
  [\href{https://arxiv.org/abs/1907.08101}{{\ttfamily 1907.08101}}].

\bibitem{Blaizot:2017ucy}
J.-P.~Blaizot and L.~Yan, \emph{{Fluid dynamics of out of equilibrium boost
  invariant plasmas}},
  \href{https://doi.org/10.1016/j.physletb.2018.02.058}{\emph{Phys. Lett. B}
  {\bfseries 780} (2018) 283}
  [\href{https://arxiv.org/abs/1712.03856}{{\ttfamily 1712.03856}}].

\bibitem{Muller:1967zza}
I.~Muller, \emph{{Zum Paradoxon der Warmeleitungstheorie}},
  \href{https://doi.org/10.1007/BF01326412}{\emph{Z. Phys.} {\bfseries 198}
  (1967) 329}.

\bibitem{Israel:1976tn}
W.~Israel, \emph{{Nonstationary irreversible thermodynamics: A Causal
  relativistic theory}},
  \href{https://doi.org/10.1016/0003-4916(76)90064-6}{\emph{Annals Phys.}
  {\bfseries 100} (1976) 310}.

\bibitem{Baier:2007ix}
R.~Baier, P.~Romatschke, D.T.~Son, A.O.~Starinets and M.A.~Stephanov,
  \emph{{Relativistic viscous hydrodynamics, conformal invariance, and
  holography}}, \href{https://doi.org/10.1088/1126-6708/2008/04/100}{\emph{J.
  High Energy Phys.} {\bfseries 04} (2008) 100}
  [\href{https://arxiv.org/abs/0712.2451}{{\ttfamily 0712.2451}}].

\bibitem{Heller:2022ejw}
M.P.~Heller, A.~Serantes, M.~Spali\'nski and B.~Withers, \emph{{Rigorous Bounds
  on Transport from Causality}},
  \href{https://doi.org/10.1103/PhysRevLett.130.261601}{\emph{Phys. Rev. Lett.}
  {\bfseries 130} (2023) 261601}
  [\href{https://arxiv.org/abs/2212.07434}{{\ttfamily 2212.07434}}].

\bibitem{Spalinski:2016fnj}
M.~Spali\'nski, \emph{{Small systems and regulator dependence in relativistic
  hydrodynamics}},
  \href{https://doi.org/10.1103/PhysRevD.94.085002}{\emph{Phys. Rev. D}
  {\bfseries 94} (2016) 085002}
  [\href{https://arxiv.org/abs/1607.06381}{{\ttfamily 1607.06381}}].

\bibitem{Florkowski:2017olj}
W.~Florkowski, M.P.~Heller and M.~Spali{\'n}ski, \emph{{New theories of
  relativistic hydrodynamics in the LHC era}},
  \href{https://doi.org/10.1088/1361-6633/aaa091}{\emph{Rept. Prog. Phys.}
  {\bfseries 81} (2018) 046001}
  [\href{https://arxiv.org/abs/1707.02282}{{\ttfamily 1707.02282}}].

\bibitem{Jankowski:2023fdz}
J.~Jankowski and M.~Spali\'nski, \emph{{Hydrodynamic attractors in
  ultrarelativistic nuclear collisions}},
  \href{https://doi.org/10.1016/j.ppnp.2023.104048}{\emph{Prog. Part. Nucl.
  Phys.} {\bfseries 132} (2023) 104048}
  [\href{https://arxiv.org/abs/2303.09414}{{\ttfamily 2303.09414}}].

\bibitem{Floerchinger:2011pxy}
S.~Floerchinger and U.A.~Wiedemann, \emph{{Fluctuations around Bjorken Flow and
  the onset of turbulent phenomena}},
  \href{https://doi.org/10.1007/JHEP11(2011)100}{\emph{JHEP} {\bfseries 11}
  (2011) 100} [\href{https://arxiv.org/abs/1108.5535}{{\ttfamily 1108.5535}}].

\bibitem{Luzum:2008cw}
M.~Luzum and P.~Romatschke, \emph{{Conformal Relativistic Viscous
  Hydrodynamics: Applications to RHIC results at s(NN)**(1/2) = 200-GeV}},
  \href{https://doi.org/10.1103/PhysRevC.78.034915}{\emph{Phys. Rev. C}
  {\bfseries 78} (2008) 034915}
  [\href{https://arxiv.org/abs/0804.4015}{{\ttfamily 0804.4015}}].

\bibitem{Price:1994pm}
R.H.~Price and J.~Pullin, \emph{{Colliding black holes: The Close limit}},
  \href{https://doi.org/10.1103/PhysRevLett.72.3297}{\emph{Phys. Rev. Lett.}
  {\bfseries 72} (1994) 3297}
  [\href{https://arxiv.org/abs/gr-qc/9402039}{{\ttfamily gr-qc/9402039}}].

\bibitem{Bhattacharyya:2007vjd}
S.~Bhattacharyya, V.E.~Hubeny, S.~Minwalla and M.~Rangamani, \emph{{Nonlinear
  Fluid Dynamics from Gravity}},
  \href{https://doi.org/10.1088/1126-6708/2008/02/045}{\emph{J. High Energy
  Phys.} {\bfseries 02} (2008) 045}
  [\href{https://arxiv.org/abs/0712.2456}{{\ttfamily 0712.2456}}].

\bibitem{Soloviev:2021lhs}
A.~Soloviev, \emph{{Hydrodynamic attractors in heavy ion collisions: a
  review}}, \href{https://doi.org/10.1140/epjc/s10052-022-10282-4}{\emph{Eur.
  Phys. J. C} {\bfseries 82} (2022) 319}
  [\href{https://arxiv.org/abs/2109.15081}{{\ttfamily 2109.15081}}].

\bibitem{Spalinski:2022cgj}
M.~Spali\'nski, \emph{{Initial State and Approach to Equilibrium}},
  \href{https://doi.org/10.5506/APhysPolBSupp.16.1-A9}{\emph{Acta Phys. Polon.
  Supp.} {\bfseries 16} (2023) 9}
  [\href{https://arxiv.org/abs/2209.13849}{{\ttfamily 2209.13849}}].

\bibitem{Aniceto:2022dnm}
I.~Aniceto, D.~Hasenbichler and A.O.~Daalhuis, \emph{{The late to early time
  behaviour of an expanding plasma: hydrodynamisation from exponential
  asymptotics}}, \href{https://doi.org/10.1088/1751-8121/acc61d}{\emph{J. Phys.
  A} {\bfseries 56} (2023) 195201}
  [\href{https://arxiv.org/abs/2207.02868}{{\ttfamily 2207.02868}}].

\bibitem{Heller:2020anv}
M.P.~Heller, R.~Jefferson, M.~Spali\'nski and V.~Svensson, \emph{{Hydrodynamic
  Attractors in Phase Space}},
  \href{https://doi.org/10.1103/PhysRevLett.125.132301}{\emph{Phys. Rev. Lett.}
  {\bfseries 125} (2020) 132301}
  [\href{https://arxiv.org/abs/2003.07368}{{\ttfamily 2003.07368}}].

\bibitem{Aniceto:2015mto}
I.~Aniceto and M.~Spali\'nski, \emph{{Resurgence in Extended Hydrodynamics}},
  \href{https://doi.org/10.1103/PhysRevD.93.085008}{\emph{Phys. Rev. D}
  {\bfseries 93} (2016) 085008}
  [\href{https://arxiv.org/abs/1511.06358}{{\ttfamily 1511.06358}}].

\bibitem{Bender78:AMM}
C.M.~Bender and S.A.~Orszag, \emph{{Advanced Mathematical Methods for
  Scientists and Engineers}}, McGraw-Hill (1978).

\bibitem{wasow1965asymptotic}
W.~Wasow, \emph{Asymptotic expansions for ordinary differential equations},
  Pure and Applied Mathematics, Vol. XIV, Interscience Publishers John Wiley \&
  Sons, Inc., New York-London-Sydney (1965).

\bibitem{PresTeukVettFlan92}
W.H.~Press, S.A.~Teukolsky, W.T.~Vetterling and B.P.~Flannery, \emph{Numerical
  Recipes in C}, Cambridge University Press, Cambridge, USA, second~ed. (1992).

\bibitem{Cooper:1974}
F.~Cooper and G.~Frye, \emph{Single-particle distribution in the hydrodynamic
  and statistical thermodynamic models of multiparticle production},
  \href{https://doi.org/10.1103/PhysRevD.10.186}{\emph{Phys. Rev. D} {\bfseries
  10} (1974) 186}.

\bibitem{Teaney:2003kp}
D.~Teaney, \emph{{The Effects of viscosity on spectra, elliptic flow, and HBT
  radii}}, \href{https://doi.org/10.1103/PhysRevC.68.034913}{\emph{Phys. Rev.
  C} {\bfseries 68} (2003) 034913}
  [\href{https://arxiv.org/abs/nucl-th/0301099}{{\ttfamily nucl-th/0301099}}].

\bibitem{Ollitrault:2023wjk}
J.-Y.~Ollitrault, \emph{{Measures of azimuthal anisotropy in high-energy
  collisions}},
  \href{https://doi.org/10.1140/epja/s10050-023-01157-7}{\emph{Eur. Phys. J. A}
  {\bfseries 59} (2023) 236}
  [\href{https://arxiv.org/abs/2308.11674}{{\ttfamily 2308.11674}}].

\bibitem{Ambrus:2022qya}
V.E.~Ambrus, S.~Schlichting and C.~Werthmann, \emph{{Establishing the Range of
  Applicability of Hydrodynamics in High-Energy Collisions}},
  \href{https://doi.org/10.1103/PhysRevLett.130.152301}{\emph{Phys. Rev. Lett.}
  {\bfseries 130} (2023) 152301}
  [\href{https://arxiv.org/abs/2211.14356}{{\ttfamily 2211.14356}}].

\bibitem{Chaudhuri_2006}
A.K.~Chaudhuri and U.~Heinz, \emph{Effects of jet quenching on the
  hydrodynamical evolution of quark-gluon plasma},
  \href{https://doi.org/10.1103/physrevlett.97.062301}{\emph{Phys. Rev. Lett.}
  {\bfseries 97} (2006) }.

\bibitem{Chesler:2007an}
P.M.~Chesler and L.G.~Yaffe, \emph{{The Wake of a quark moving through a
  strongly-coupled plasma}},
  \href{https://doi.org/10.1103/PhysRevLett.99.152001}{\emph{Phys. Rev. Lett.}
  {\bfseries 99} (2007) 152001}
  [\href{https://arxiv.org/abs/0706.0368}{{\ttfamily 0706.0368}}].

\bibitem{Casalderrey-Solana:2020rsj}
J.~Casalderrey-Solana, J.G.~Milhano, D.~Pablos, K.~Rajagopal and X.~Yao,
  \emph{{Jet Wake from Linearized Hydrodynamics}},
  \href{https://doi.org/10.1007/JHEP05(2021)230}{\emph{J. High Energy Phys.}
  {\bfseries 05} (2021) 230}
  [\href{https://arxiv.org/abs/2010.01140}{{\ttfamily 2010.01140}}].

\bibitem{Yang_2023}
Z.~Yang, T.~Luo, W.~Chen, L.~Pang and X.-N.~Wang, \emph{3d structure of
  jet-induced diffusion wake in an expanding quark-gluon plasma},
  \href{https://doi.org/10.1103/physrevlett.130.052301}{\emph{Phys. Rev. Lett.}
  {\bfseries 130} (2023) }.

\bibitem{Akamatsu:2017}
Y.~Akamatsu, A.~Mazeliauskas and D.~Teaney, \emph{{A kinetic regime of
  hydrodynamic fluctuations and long time tails for a Bjorken expansion}},
  \href{https://doi.org/10.1103/PhysRevC.95.014909}{\emph{Phys. Rev. C}
  {\bfseries 95} (2017) 014909}
  [\href{https://arxiv.org/abs/1606.07742}{{\ttfamily 1606.07742}}].

\bibitem{An:2019rhf}
X.~An, G.~Ba\c{s}ar, M.~Stephanov and H.-U.~Yee, \emph{{Relativistic
  Hydrodynamic Fluctuations}},
  \href{https://doi.org/10.1103/PhysRevC.100.024910}{\emph{Phys. Rev. C}
  {\bfseries 100} (2019) 024910}
  [\href{https://arxiv.org/abs/1902.09517}{{\ttfamily 1902.09517}}].

\bibitem{An:2019fdc}
X.~An, G.~Ba\c{s}ar, M.~Stephanov and H.-U.~Yee, \emph{{Fluctuation dynamics in
  a relativistic fluid with a critical point}},
  \href{https://doi.org/10.1103/PhysRevC.102.034901}{\emph{Phys. Rev. C}
  {\bfseries 102} (2020) 034901}
  [\href{https://arxiv.org/abs/1912.13456}{{\ttfamily 1912.13456}}].

\bibitem{An_2021}
X.~An, G.~Ba{\c{s}}ar, M.~Stephanov and H.-U.~Yee, \emph{Evolution of
  non-gaussian hydrodynamic fluctuations},
  \href{https://doi.org/10.1103/PhysRevLett.127.072301}{\emph{Phys. Rev. Lett.}
  {\bfseries 127} (2021) 072301}
  [\href{https://arxiv.org/abs/2009.10742}{{\ttfamily 2009.10742}}].

\bibitem{An_2023}
X.~An, G.~Ba\c{s}ar, M.~Stephanov and H.-U.~Yee, \emph{{Non-Gaussian
  fluctuation dynamics in relativistic fluids}},
  \href{https://doi.org/10.1103/PhysRevC.108.034910}{\emph{Phys. Rev. C}
  {\bfseries 108} (2023) 034910}
  [\href{https://arxiv.org/abs/2212.14029}{{\ttfamily 2212.14029}}].

\bibitem{Chen:2022ryi}
Z.~Chen, D.~Teaney and L.~Yan, \emph{{Hydrodynamic attractor of noisy
  plasmas}}, \href{https://doi.org/10.1103/PhysRevC.108.064911}{\emph{Phys.
  Rev. C} {\bfseries 108} (2023) 064911}
  [\href{https://arxiv.org/abs/2206.12778}{{\ttfamily 2206.12778}}].

\end{thebibliography}\endgroup

\end{document}